\documentclass[fleqn,usenatbib]{mnras}

\usepackage[T1]{fontenc}

\DeclareRobustCommand{\VAN}[3]{#2}
\let\VANthebibliography\thebibliography
\def\thebibliography{\DeclareRobustCommand{\VAN}[3]{##3}\VANthebibliography}

\usepackage{graphicx}	
\usepackage{amsmath}	
\usepackage{amssymb}	
\usepackage{txfonts}
\usepackage{indentfirst}
\usepackage{ulem}
\usepackage{listings}

\title[CoBiToM Project - I. Contact Binaries Towards Merging]{CoBiToM Project - I. Contact Binaries Towards Merging}

\author[K. Gazeas et al.]{
K. D. Gazeas,$^{1}$\thanks{E-mail: kgaze@phys.uoa.gr}
G. A. Loukaidou,$^{1}$
P. G. Niarchos,$^{1}$
S. Palafouta,$^{1}$
D. Athanasopoulos,$^{1}$
\newauthor
A. Liakos,$^{2}$
S. Zola,$^{3,4}$
A. Essam$^{5}$
and P. Hakala$^{6}$
\\
$^{1}$ Section of Astrophysics, Astronomy and Mechanics, Department of Physics, National and Kapodistrian University of Athens, \\ GR-15784 Zografos, Athens, Greece\\
$^{2}$ IAASARS, National Observatory of Athens, GR-15236 Penteli, Athens, Greece\\
$^{3}$ Astronomical Observatory, Jagiellonian University, ul. Orla 171, PL-30-244 Krakow, Poland\\
$^{4}$ Mt. Suhora Observatory, Pedagogical University, ul. Podchorazych 2, PL-30-084 Krakow, Poland\\
$^{5}$ Department of Astronomy, National Research Institute of Astronomy and Geophysics (NRIAG), Helwan, Cairo 11421, Egypt\\
$^{6}$ Finnish Centre for Astronomy with ESO (FINCA), Quantum, University of Turku, FI-20014, Turku, Finland
}

\date{Accepted XXX. Received YYY; in original form ZZZ}

\pubyear{2021}

\begin{document}
\label{firstpage}
\pagerange{\pageref{firstpage}--\pageref{lastpage}}
\maketitle

\begin{abstract}
Binary and multiple stellar systems are numerous in our solar neighborhood with
80~per~cent of the solar-type stars being members of systems with high order multiplicity.
The $Contact~Binaries~Towards~Merging$ {\it (CoBiToM) Project} is a programme that
focuses on contact binaries and multiple stellar systems, as a key for
understanding stellar nature. The goal is to investigate stellar
coalescence and merging processes, as the final state of stellar evolution of low-mass contact binary systems.
Obtaining observational data of approximately
100 eclipsing binaries and multiple systems and more than 400 archival systems, the
programme aspires to give insights for their physical and orbital parameters
and their temporal variations, e.g. the orbital period modulation, spot activity
etc. Gravitational phenomena in multiple-star environments will be linked
with stellar evolution.
A comprehensive analysis will be conducted, in order to investigate the
possibility of contact binaries to host planets, as well as the link between
inflated hot Jupiters and stellar mergers. The innovation of {\it CoBiToM Project}
is based on a multi-method approach and a detailed investigation, that will
shed light for the first time on the origin of stellar mergers and rapidly
rotating stars.
In this work we describe the scientific rationale, the observing facilities to be used and the methods
that will be followed to achieve the goals of {\it CoBiToM Project} and we present the first results as an example of the current research on evolution of contact binary systems.

\end{abstract}



\section{Scientific Rationale}

Contact binaries are the most frequently observed type of eclipsing binary systems. Their components are small in size in comparison to other Main Sequence stars, they are cool, low-mass, and they belong to the old stellar population group \citep{Kaluzny1993,Rucinski1998,Rucinski2000,Rucinski2013}. They follow certain empirical relationships, which closely correlate physical parameters with each other, being a consequence of their configuration within the Roche geometry
\citep{Gazeas2009, Stepien2012, yan2015}.

Contact binaries are excellent astrophysical tools for the determination of the fundamental physical parameters (mass, radius, luminosity) of low-temperature stellar components, that can be used to trace their evolutionary paths. They are of great interest in the study of stellar populations since they can be used as distance indicators \citep{Rucinski&Duerberk1997, Rubenstein1996, Edmonds1996, Yan1993}. Stellar evolution studies have been mainly performed by means of photometry and spectroscopy. As a result, contact binaries provide an excellent opportunity to investigate stellar merging scenarios.

Stellar evolution theories indicate that we should be able to observe stellar merging events \citep{Webbink1976,Stepien2006a}
due to the gradual coalescence of contact binaries. However, this has not been the case until the detection of the merging event of V1309~Sco in 2008 \citep{Tylenda2011}. Even in this case, the contact binary nature of the progenitor was discovered only after the red nova event (Nova Sco 2008). Numerous observations of V1309~Sco are available in the
OGLE-III and IV surveys \citep{Udalski2003} prior to the red nova event and they helped to identify the nova progenitor as a contact binary with an orbital period of $\sim1.4$~d. \cite{Kochanek2014} estimated that stellar mergers in our Galaxy can occur approximately once every 10 years. There are contact binaries that recently were verified as stellar merger candidates such as ZZ~PsA \citep{Wadhwa2020}, while more cases are waiting to be confirmed from databases and catalogues. FK~Com, HD~199178, and UZ~Lib are single and rapidly rotating stars and may also be interpreted as the result of a short-orbital period binary merger \citep{Stepien1995,Mink2011}. Understanding short-orbital period contact binaries and rapidly rotating stars, will hopefully link stellar mergers with Blue Stragglers, which are observed in evolved clusters.
{\it CoBiToM Project} is a new observing programme, which was initiated at the University of Athens in 2012, aiming to investigate the possible lower limit of the orbital period of contact binary systems before coalescence as a prediction tool of stellar evolution prior to merging.

Binary and multiple stellar systems populate the entire Galaxy. \cite{Duchene2013} indicate a multiplicity fraction for intermediate-mass stars to be more than 50~per~cent, rising to more than 80~per~cent for the most massive stars. Apparently, single stars appear to be the minority, especially in the higher mass regime. Stellar evolution theories strongly suggest that binary interactions are probably responsible for creating several different types of supernovae, novae and unusual star types such as Blue Stragglers.

Eclipsing binaries can be also found in clusters, as well as among field stars. There is an extensive study of such systems, as members of evolved globular clusters \citep{Kaluzny1990, Kaluzny1992, Kaluzny1993, Hut1992}, the majority of which are in contact configuration. An open question still exists on whether stellar mergers are connected with other stellar populations such as the newly discovered eclipsing binaries of R~CMa type \citep{Budding2011}, fast rotators and Blue Stragglers. {\it CoBiToM Project} will make an effort to link these (phenomenologically separate) groups together and explain stellar evolution in cluster environments, as discussed by \cite{Parker2009}. The investigation of inflated hot Jupiters will constrain the possible origins of giant planets and indicate a potential search area for exoplanet hunters.

Contact binaries evolve very slowly in comparison to the more massive stars, a process which is usually accompanied with mass and angular momentum loss. As a result, they have short orbital period and low angular momentum. This configuration eventually leads towards merging, but at a rather slow pace. There are also strong arguments implying a merging process \citep[e.g.][]{Pietrukowicz2017,Wadhwa2020}, such as the rapid orbital period change, the decreasing separation between the components, the non-synchronous rotation, as well as the angular momentum and mass loss from the system. The main goal of {\it CoBiToM Project} is to decipher how the processes towards merging may occur, by investigating the changes in the physical parameters of the members of the systems. Such an approach has never been implemented up to date, therefore the results of this project will be pioneering and will contribute to a significant improvement in our knowledge for the merging processes.

Although low-mass dwarf stars are very common, their evolution in contact binary systems is poorly understood. It has been clear for a number of years that there is a short period cut-off for such systems around an orbital period of about 0.22 d \citep{Rucinski2007}. The reason for this period cut-off is now attributed to magnetic wind-driven angular momentum loss mechanisms, and, hence, linked to the finite age of the binary population \citep{Stepien2006a}. Detailed calculations by the same author showed that the time needed to reach the stage of Roche lobe overflow (RLOF) and, hence, to appear as a contact binary, is around 7.5~Gyr for a primary star of 1~$M_{\sun}$. However, because the angular momentum loss timescale increases substantially with decreasing stellar mass, the time to reach RLOF for a system with a primary star of mass 0.7 $M_{\odot}$ increases to a value greater than the age of the Universe. The short period cut-off for contact binaries is therefore suggested to be due to the finite age of the Galaxy, i.e. lower mass stars in binaries which could evolve to shorter period systems have not yet had enough time to do so. An alternative theory \citep{Molnar2017} suggests that the cut-off is a consequence of higher apsidal precession rate in smaller primaries that prevents formation of contact binaries via Kozai cycles.

\begin{figure*}
\includegraphics[width=5.8cm,scale=1.0,angle=0]{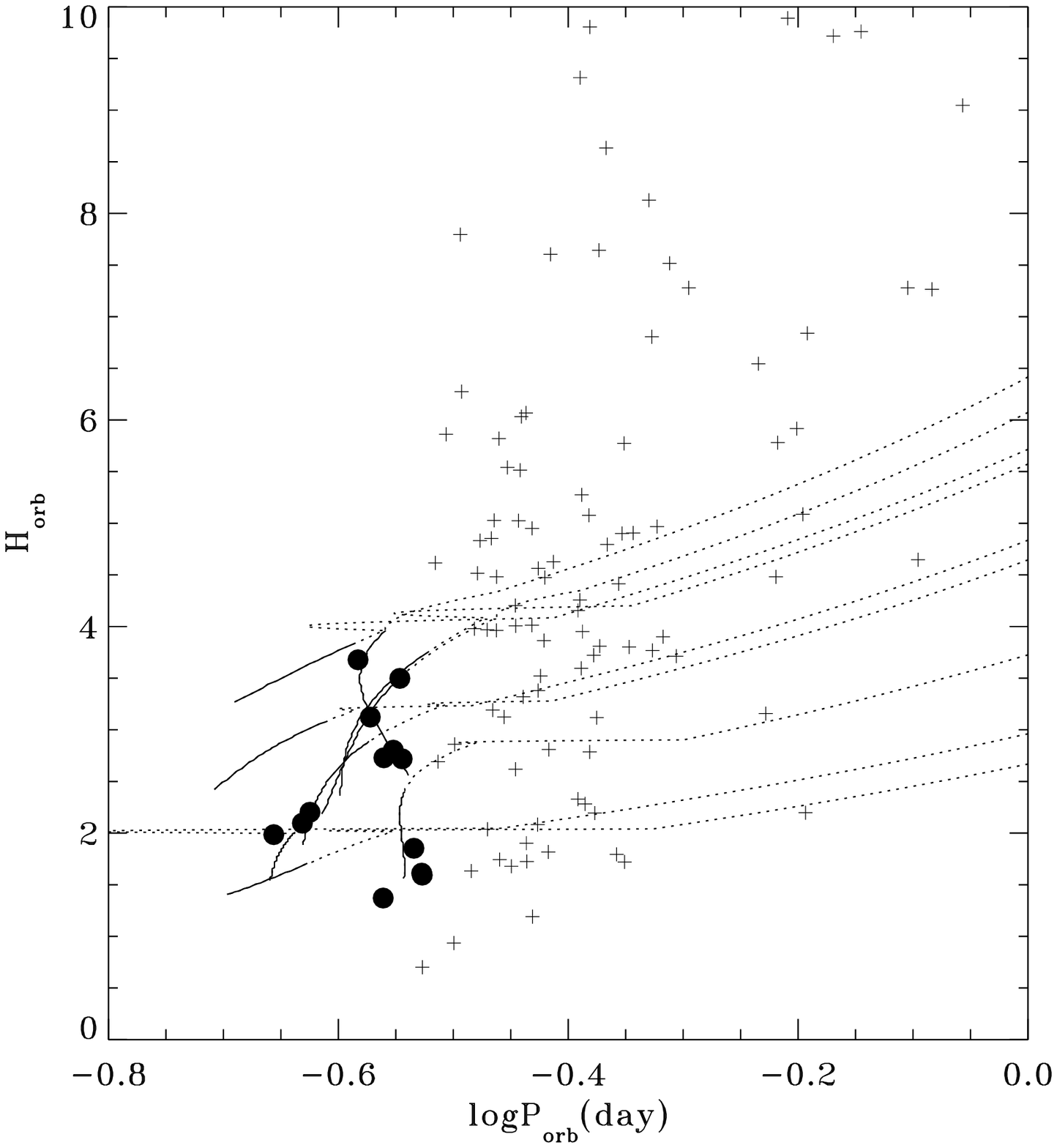}
\includegraphics[width=5.8cm,scale=1.0,angle=0]{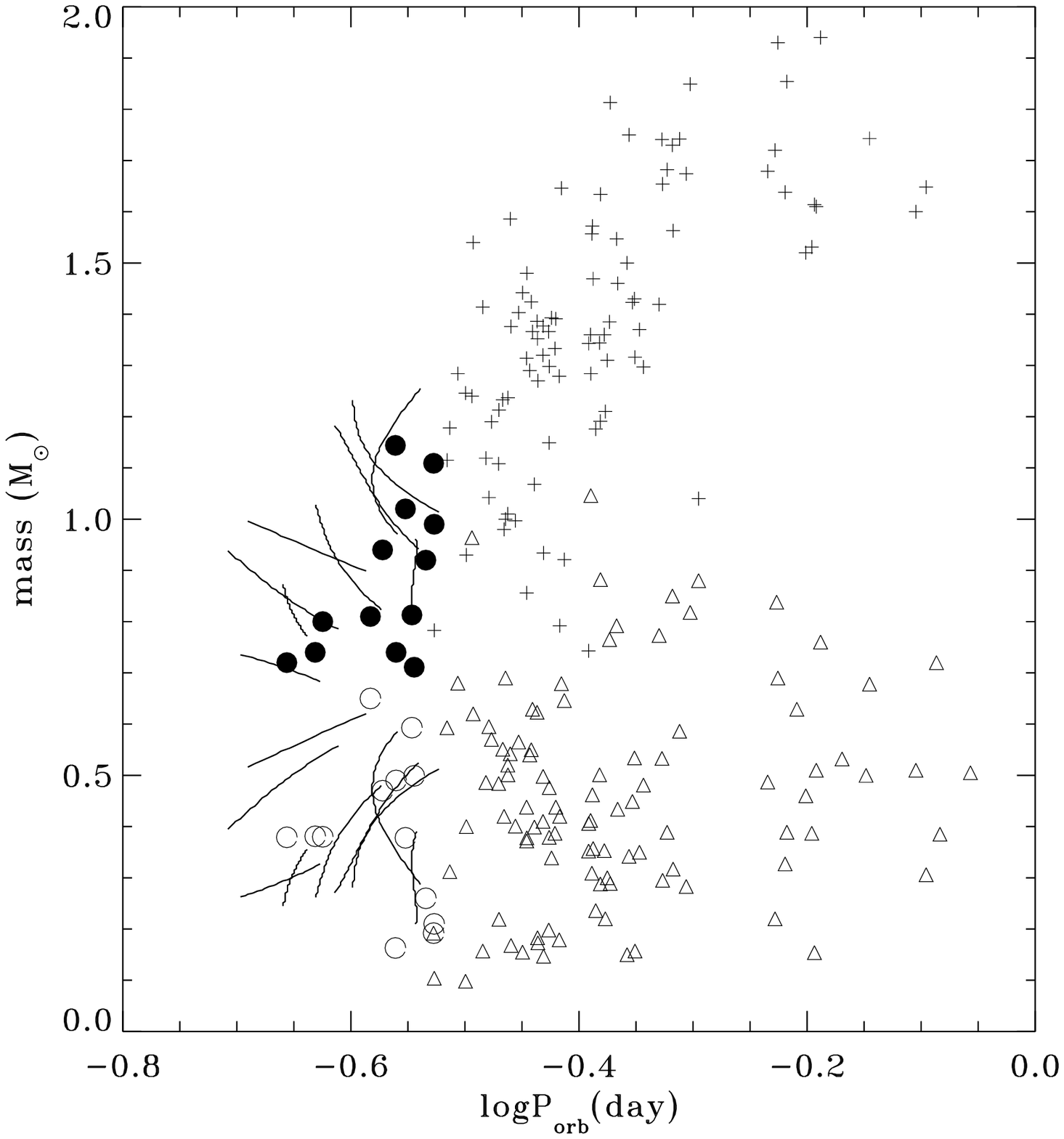}
\includegraphics[width=5.8cm,scale=1.0,angle=0]{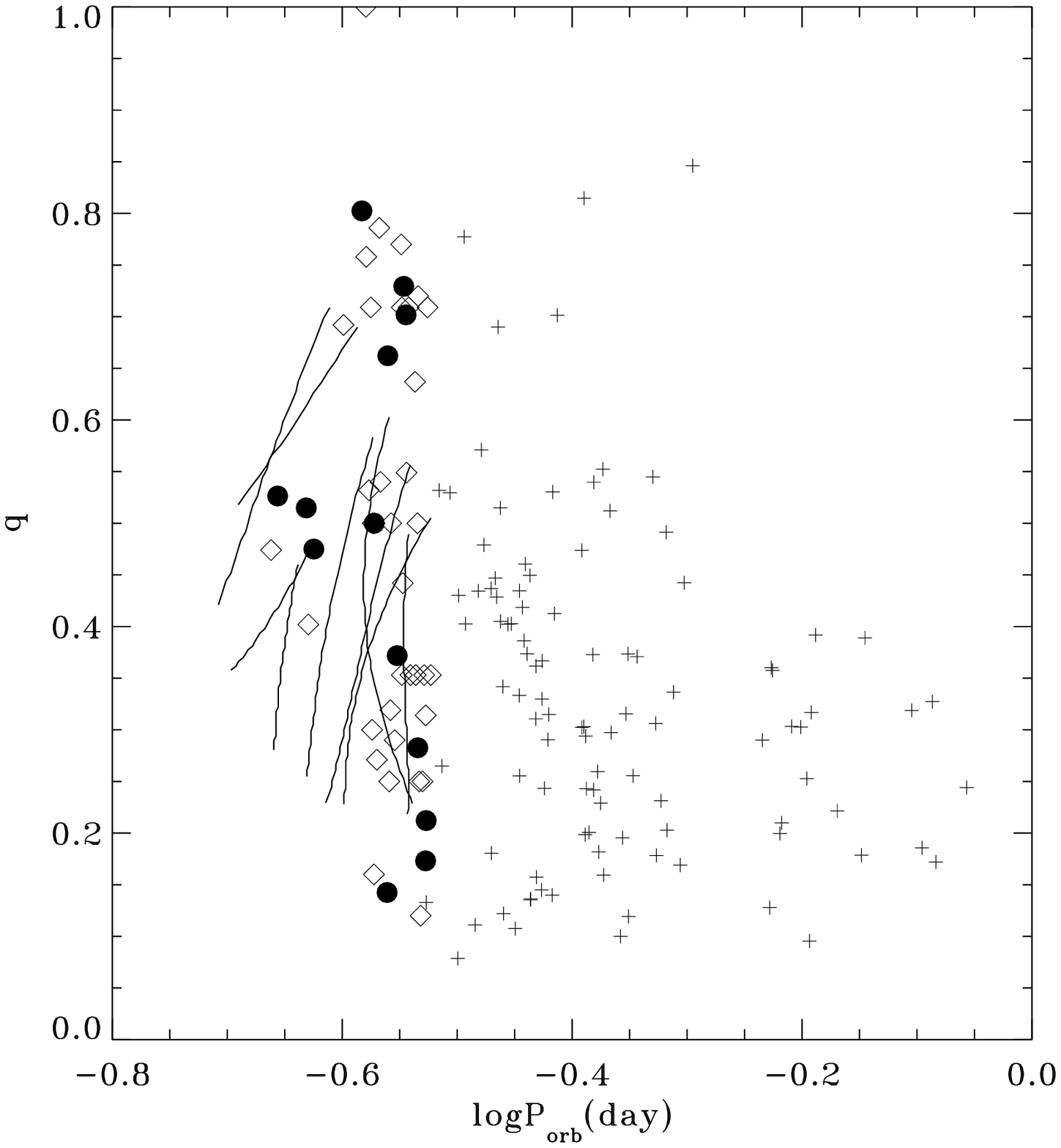}\\
\caption{Left panel: The distribution of angular momentum of 112 contact binaries in cgs units 
($\times10^{51}$). Evolutionary tracks of models given by \protect\cite{Stepien2012} are also 
shown. Parts of the tracks plotted with dotted lines correspond to pre-contact phases 
and those plotted with solid lines describe binary evolution while in contact configuration.
Middle panel: The components of low-mass contact binaries (open and filled circles) 
and of other contact binary systems (plus signs and triangles). These values are 
compared with the models of  contact binaries \protect\citep{Gazeas2008}.
Right panel: The spectroscopically determined mass ratio of the same contact binaries, 
supplemented by the photometrically determined mass ratio of the low-mass contact 
binaries (open and filled circles) and other W~UMa-type systems (plus signs). 
They are compared with the evolutionary models (solid lines) for a sample of binaries 
that evolve from a detached configuration towards contact one and eventually merging.
}
\label{Fig1}
\end{figure*}

\cite{Gazeas2008} demonstrated that there are clear correlations between
the mass of the components of contact binaries and their orbital period, as well as between the radius of the components and their orbital period (Fig. \ref{Fig1}). In particular, the period cut-off of around 0.22~d corresponds to primary and secondary mass of around 0.85~$M_{\sun}$ and 0.3~$M_{\sun}$ and radius of around 0.7~$R_{\sun}$ and 0.5~$R_{\sun}$, respectively. The short period end of these correlations is defined by only a few systems that present considerable scatter among their parameters. Fig. \ref{Fig1} presents the distribution of contact binaries with respect to their orbital period. The left panel shows that the angular momentum decreases as the binary system evolves. The middle panel shows that the orbital period changes during the mass transfer process. The right panel shows that the mass ratio is reduced, as the orbit shrinks and the merging process occurs. Increasing the number of systems in this period range will allow these correlations to be further tested and improved upon.

The structure of the current work is as follows: After a brief overview and scientific rationale of the {\it CoBiToM Project} given in Introduction, a detailed description follows in section 2, presenting the expected results. The observing strategy, data reduction and modeling are described in the following two sections (3 and 4), explaining the data gathering process and methods of analysis. First results for a selected sample from the {\it CoBiToM Project} are given in section 5, showing the potential of the project to perform the expected study and fulfill the goal stated initially. The discussion summarizes the major findings of this study, encourage future studies of stellar evolution and highlights the importance and the scientific impact of the research in the field of contact binaries in Astrophysics.


\section{Methodology and Expected Results}

In the analysis of W~UMa stars, the light curves are phased with reference to the deeper minimum. The star eclipsed at phase = 0.0 is called the `primary component', while the `secondary component' is eclipsed at phase = 0.5.
This logically implies that the `primary' is the more massive and hotter between the two components,
which is not always the case in binary systems. Spectroscopy can determine the more massive from the less massive component and only the combination of photometry and spectroscopy together provides a clear picture of the system and the mass-luminosity relation. In this study we define as `primary' the more massive component, in order to be consistent with the definition of `primary' and `secondary' terms in spectroscopy. Therefore, the mass ratio, defined as $q~=~M_{2}/M_{1}$, will be always less than unity.

{\it CoBiToM Project} grounds its originality in the concurrent examination of the merging processes from different perspectives. The idea of the project is based on a holistic approach consisting of:

\begin{enumerate}
\item Determination of absolute physical parameters of a large sample of binaries (> 50 systems) by utilizing space-borne data, performing ground-based follow-up photometric and spectroscopic observations and inspecting the temporal variations of orbital and physical parameters.

\item Expansion and reassessment of the sample of low-mass, short-period contact binaries and stellar merger candidates that exhibit large orbital period change rate ($dP/dt$) and detailed study of their evolution state as a probe for stellar merging scenarios.
All available literature data are being gathered for long-term $Observed-Calculated$ ($O-C$) diagrams, i.e. {\it Eclipse Timing Variations} (ETV). This will provide a clear picture of orbital period modulation mechanisms (e.g. mass transfer, third body orbiting the system etc.).
Studies based on SWASP \citep{Lohr2012}, OGLE \citep{Kubiak2006}, and {\it Kepler} \citep{Conroy2014} data are a few examples of such statistical analysis of large sample of contact binaries.
Most of them, however, cover a short time span, making it difficult to study accurately the orbital period modulations and only a few of them focus specifically on low-mass, low-q or short-period contact binaries.

\item Investigation of multiple system environments in the context of how the Kozai cycle mechanism \citep{Kozai1962}
affects close and apparently contact binary formation and drives the corresponding evolution channels.
An extensive research on this field has been made by \cite{Zasche2009a}, who pointed out the importance of studying a multiple systems as a probe to disentangle the mechanisms of multiple systems formation and evolution.

\item Search for a possible link between short-orbital period contact binaries and other stellar populations, such as the fast rotators \citep[i.e. FK~Com, Blue Stragglers;][]{Stepien1995,Knigge2009} or even with the unusual Algol systems of R~CMa type that exhibit low mass ratio and short orbital period \citep{Budding2011}. Our findings are expected to provide concrete evidence of an existing link between stellar merger candidates through stellar collisions in evolved clusters.

\item Delineation of the planetary stability zone for each binary system and a possible relation between host stars of inflated hot Jupiters and stellar mergers. This study aspires to reveal, for the first time, how the evolution of the components in a binary system will affect the stability zone. Numerical simulations will be performed, in order to examine the possibility for a binary to form or host a planet after a coalescence event, as suggested by \citet{Tutukov1991,Tutukov2004} and
\citet{Martin2011}.

\end{enumerate}


Ground-based sky surveys such as SuperWASP, OGLE, ASAS, 2MASS, CSS, RATS \citep[respectively]{Pollacco2006,Soszynski2015,Paczynski2006,Gettel2005,Drake2014,Ramsay2005} have already identified approximately 400 contact binaries (with short orbital period and total eclipses), which will be studied in this project. Space-borne data from {\it Kepler} and {\it CoRoT} missions \citep[respectively]{Prsa2011, Deleuil2018} have added 20 more such systems, that fulfill the same criteria. All these systems will be used to derive a reliable conclusion for the major merging aspects, to examine the pre-merging state and to constrain the timescale of stellar merging and evolution.
The updated Multiple Star Catalogue \citep[MSC;][]{Tokovinin2018} includes about 2000 multiple systems, with the majority being hierarchical. However, quadruple systems that are gravitationally bound and produce mutual eclipses are very rare. Only 28 doubly eclipsing systems show strong evidence of relative motions. In addition, the narrow sky area around the Large Magellanic Cloud, covered by the OGLE survey, provides 146 more candidates \citep{Zasche2019}, awaiting to be confirmed by ground-based observations. It is expected that at least 20 multiple systems from the aforementioned sample will be studied in the frame of this project, making a significant contribution to the study of multiple systems and the orbital evolution processes. The systems of special interest (i.e. those with extremely short orbital period, low mass ratio) will only be observed in follow-up mode in order to monitor accurately their light curve variability ($<$10~mmag) and radial velocity values ($<$5~km~s$^{-1}$).

In addition, more than 3000 stars have been found to host exoplanets, 60~per~cent of which ($\sim2000$) are solar-type or cooler stars and host hot Jupiters with known mass ({\it `NASA Exoplanet Archive'}\footnote{\url{https://exoplanetarchive.ipac.caltech.edu/}}). Orbital and physical properties of these exoplanets will be examined along with the characteristics of their host stars. It is expected that correlations between the orbital characteristics of hot Jupiters with inflated radius and the properties of their host stars will be found, revealing whether there is a possible link with stellar mergers.

This study will lead to solid results in the following topics:

\begin{enumerate}
\item Orbital shrinking: Merging scenarios and their corresponding parameters are not constrained sufficiently. {\it CoBiToM Project} will include orbital period analysis of all systems under investigation. Systems with negative $dP/dt$ value indicate that they exhibit orbital shrinking. Cyclic modulation of orbital period can be connected either with the Applegate mechanism \citep{Applegate1992} and, hence, magnetic braking or with the existence of a third body orbiting the eclipsing systems, which will further affect the period modulation and favor Kozai cycles. The apsidal motion of an eccentric orbit can also produce orbital period modulation, however, this does not apply on the circular orbits of contact binaries. An extended list of stellar merger candidates will be created using all available ground-based and space-borne data.
	
\item Evolution state of stellar systems: Stellar evolution of binary and multiple systems is totally different from that of single stars and this is shown in all H-R diagrams (or related $M-R$, $M-L$ diagrams), where the components obviously do not follow the same trend as the single Main Sequence stars do. This is a result of the location of stellar components with respect to the ZAMS and TAMS limits. The secondary components of binary systems seem to be oversized for their mass, which lead them to become larger, in order to preserve stability.  The {\it CoBiToM Project} aims to show that this process also applies to ultra-short period systems and that it is a result of stellar evolution.

\item Mutually eclipsing multiple systems: Such a configuration is very rare as the complexity increases significantly with every additional component \citep{Tokovinin2018a}. Quadruple systems are still an intriguing question, with a possibility that various possible hierarchical configurations
are formed via different evolution channels \citep{Tokovinin2018b}. The nature of quadruple and multiple systems favors coalescence or orbital shrinkage via Kozai cycles \citep{Kozai1962}, faster than in a `naked' (isolated) binary system. About 70~per~cent of contact binary systems have additional components in a close or wide orbit, being triples or multiples \citep{Pribulla2008,Deb2011,D'Angelo2006}. Apparently, angular momentum loss from the inner close binaries are favored via Kozai cycles.

\item Darwin instability test probes: In order to probe Darwin instability \citep{Darwin1879} in contact binaries, the absolute physical parameters and orbital angular momenta have to be determined. Orbital angular momentum loss (AML), accompanied with a simultaneous increase of the primary component’s spin, is an indication that the orbital angular momentum is absorbed by the spinning-up of the primary. This case can ensure that these systems are heading into a very unstable phase and can be characterized as stellar merger candidates.

\item Mergers are a possible host place for planetary formation: Contact binaries that are driven into coalescence could possibly host a circumbinary disk from the lost matter from which planets may form \citep{Tutukov1991,Tutukov2004}. Hundreds of hot Jupiters have been discovered, but the inflated radius of a large fraction of them remains unexplained. There is a (strong) hypothesis that hot Jupiters with inflated size originated in a disk formed by  merging of two low-mass stars \citep{Martin2011}. The nature and abundance of contact binaries of W UMa-type make them a plausible candidate population for forming hot Jupiters. The distribution of the (total) mass of contact binaries in the catalogue of \cite{Gazeas2008} peaks at 1.5-2.0 $M_{\sun}$, while the mass distribution of stars hosting transiting planets around 1.1 $M_{\sun}$ (Fig. 2 in that work). It is clear that the overlap between them is in fact only modest. Numerical simulations will be performed, in order to investigate the origin of hot Jupiters via  formation at a distance followed by inward orbital migration.

\end{enumerate}


\section{The observing strategy}

{\it CoBiToM Project} will provide long-term and continuous monitoring of contact binaries from different ground locations.
The data acquisition provides multi-color imaging in $BVRI$ bands, using one site and one instrument for each target, in order to avoid any systematic effect between instrumental calibration and/or spectral (filter) mismatch. The photometric data will be converted into normalized flux for direct comparison with the theoretical models.
Along with high accurate space-borne data from various past and present missions (i.e. Kepler, CoRoT, Gaia, TESS), excellent sets of combined photometric and spectroscopic data will be gathered.

The majority of studies in contact binaries (and eclipsing binaries in general) are based only on photometric observations \citep{Paczynski2006, Norton2011, Pietrukowicz2013, Pawlak2016, Soszynski2016, Li2019, Zhang2019, Li2020, qian2020}.
Superior accuracy on stellar parameter determination is clearly obtained when combined photometric and spectroscopic studies are performed \citep[e.g. the $W~UMa~Programme$;][]{Kreiner2006}. Therefore, spectroscopic observations will be included in our analysis. In parallel, the solutions on individual systems from photometric surveys have to be carefully examined. Nevertheless, the statistical value of a large sample of systems is always highly important.

The data from ground-based sky-surveys, space missions and those that will be obtained from the {\it CoBiToM Project} collaborating telescopes will be used for studying the following topics:

\begin{enumerate}

\item Contact Binaries: We plan to gather more than 100 contact binaries of W~UMa type, with short orbital period ($P<0.3$ d), or low mass ratio ($q<0.1$) or  period change rate ($|dP/dt|$ > 10$^{-8}$~d~yr$^{-1}$). Targets of interest and with extreme parameter values will be prioritized for observations. This sample will be created with the available space-borne and ground-based survey data.

\item Multiple systems: At least 20 multiple systems will be studied, using ground-based and space-borne data. Each selected system will include at least one contact binary and will have confirmed gravitational bound between its components. Observations of such targets is a significantly time consuming process, since they exhibit longer orbital period than the contact systems, usually of the order of 5-10 days.

\item Hot Jupiter candidates: The sample of targets will be defined and selected among solar type (G-type) stars or even late-type stars in order to be consistent with the spectral type of the cool components of evolved contact binaries. An additional criterion is the radius of hot Jupiters to be inflated. Their publicly available data will be collected from various observational campaigns and space missions, through online databases like {\it `NASA Exoplanet Archive'} \citep{Akeson2011}, {\it `Extrasolar Planets Encyclopaedia'} \citep{Schneider2011}, and {\it `Exoplanet Orbit Database'} \citep{Han2014}.

\end{enumerate}

The following observatories will be used to implement the observing project:

\begin{enumerate}
\item {\bf University of Athens Observatory - UOAO\footnote{\url{http://observatory.phys.uoa.gr}}} \\
The University of Athens Observatory (UOAO) belongs to the National and Kapodistrian University of Athens in Greece. It utilizes a 0.4~m (f/8) Cassegrain telescope, which is equipped with a SBIG~ST-10~XME CCD detector and a set of $UBVRI$ filters (Bessell specifications) in order to perform multi-band photometry. The alternative use of a f/6.3 focal reducer increases the field of view when necessary. The telescope operates as a robotic instrument and it can be remotely controlled since 2012 \citep{Gazeas2016}. A medium-resolution stellar spectrograph (UOAO-MRS) operates at a resolving power of $R$=10000 for point sources down to 7~mag.

\item {\bf Helmos Observatory}\footnote{\url{http://helmos.astro.noa.gr}} \\
Helmos observatory is operated by the National Observatory of Athens. It utilizes a 2.3 m (f/8) Ritchey- Chr\'{e}tien telescope (`Aristarchos'). The telescope may operate in various imaging modes. The first includes the Princeton Instruments VersArray 2048B LN CCD camera and a set of $UBVRI$ filters.
The second includes the RISE2 instrument that consists of an Andor CCD camera (sensor: E2V CCD47-20) and a broad $VR$ photometric filter.
Within 2021, a new Andor IKON-L CCD camera is planned to be installed and replace the LN CCD. The {\it Aristarchos Transient Spectrometer} \citep[ATS;][]{Boumis2004} is a low dispersion spectrometer (600~lines~mm$^{-1}$) is attached on this telescope. The spectrometer utilizes an Apogee U47-MB, E2V-CCD4710 AIMO back illuminated, Grade 1 CCD camera with 1k$\times$1k pixels and provides a resolution of $\sim$3.2 \AA~pixel$^{-1}$ and a spectral coverage between approximately 4000-7260 \AA.

\item {\bf Kryoneri Astronomical Station\footnote{\url{http://kryoneri.astro.noa.gr}}} \\
Kryoneri observatory is also operated by the National Observatory of Athens and utilizes a 1.2 m (f/3) prime focus telescope \citep{Xilouris2018}. Two imaging modes are available for this telescope. The first includes the twin fast frame camera system (ANDOR Zyla 5.5 sCMOS) on a beam splitter equipped with the $R$ and $I$ filters, respectively. This setup has an f/2.8 effective focal ratio.
The second option concerns the Apogee ASPEN CG-47 CCD camera and one filter from the $UBVRI$ set.
In the following years, the new Andor IKON-L CCD camera and a filter box with the $UBVRI$ filters set is planned to be installed and replace the ASPEN CCD.

\item {\bf Kottamia Astronomical Observatory - KAO\footnote{\url{http://www.nriag.sci.eg}}}\\
Kottamia Astronomical Observatory (KAO) was formerly known as Helwan Observatory. It belongs to the Department of Astronomy of the National Research Institute of Astronomy and Geophysics (NRIAG), in Egypt. It utilizes a 1.88~m telescope that operates in Newtonian (f/4.9) and Cassegrain (f/18) foci. An upgrade of the optical system and a refurbishment of the entire telescope control system was implemented in 2008 \citep{Azzam2014}, enabling computer control of both dome and telescope. The recently developed ‘Kottamia Faint Imaging Spectro-Polarimeter’ (KFISP) operates as a multi-purpose Cassegrain instrument, that incorporates a focal reducer and creates an effective focal ratio of f/6. The KFISP can be used for direct imaging, spectroscopy (low and medium resolution), polarimetric imaging, and spectro-polarimetric measurements \citep{Azzam2020}

\item {\bf Jagiellonian University Astronomical Observatory\footnote{\url{http://www.oa.uj.edu.pl}}}\\
The Jagiellonian University Astronomical Observatory is located in Krakow, Poland. It utilizes a 0.5~m (f/15) Cassegrain telescope and a 0.5~m CDK500 (f/6.8) corrected Dall-Kirkham astrograph telescope. The observatory is equipped with three back illuminated CCD cameras: the Alta F42 with an E2V CCD42-40 chip (2k$\times$2k~pixels), the Alta U42 with an E2V CCD42-40 chip (2k$\times$2k~pixels) and the Alta U47 with an E2V CCD47-10 chip (1k$\times$1k~pixels). The cameras are combined with separate sets of wide band Bessell/Johnson-Morgan filters.

\item {\bf Mt. Suhora Astronomical Observatory\footnote{\url{https://www.as.up.krakow.pl}}}\\
Mt. Suhora Astronomical Observatory belongs to the Astronomy Department of the Pedagogical University of Krakow, Poland. It utilizes a 0.6~m (f/12.5) Cassegrain telescope with the Apogee ASPEN CG-47 CCD camera 
and sets of Johnson-Morgan, Sloan and Str{\"o}mgren filters. The instrumentation works in prime focus with a  focal ratio of  f/4.
\end{enumerate}

Additional photometric information will be provided by the space-borne missions, which run for a certain period of time or they are currently operational. These missions provide highly accurate photometric data but for a limited duration. Photometric information will be also collected from several sky survey programmes, such as: SuperWASP, OGLE, ASAS, 2MASS, CSS etc, that can provide data for longer period of time.

Spectroscopic radial velocity data provide the best measurements of the system mass-ratio. These data will be obtained using the ESO telescopes for southern targets and the 2.56~m Nordic Optical Telescope (NOT) on La Palma for the northern targets through the international collaboration and ESO membership of the involved partners to {\it CoBiToM Project}.


\section{Data Reduction and Modelling}

A combined spectroscopic and photometric orbital solution will be our main methodology of data analysis and modeling, in order to determine the stellar parameters of both components. This will utilize the widely used Wilson-Devinney (W-D) code \citep{WilsonDevinney1971, Wilson1990} appended with the Monte-Carlo (MC) algorithm \citep{Zola1997} or with the Bayesian MCMC sampling methods as the search procedure in the parameter space \citep[{\it PHOEBE}\footnote{\url{http://phoebe-project.org}} software,][]{Conroy2020}.

In most studies the majority of contact binaries are observed only photometrically and the mass ratio is estimated via the so-called $q$-search method. This technique is widely used when no radial velocity measurements are available, in both contact binaries \citep[e.g.][]{Niarchos1994, Wadhwa2020} or detached systems \citep[e.g.][]{Liakos2017}. The basic idea of this method is that the mass ratio is excluded from the list of free parameters. The modelling code (in this case the W-D code) produces, for the given $q$ value, a sum of squared residuals ($\Sigma res^2$). This procedure is repeated for a range of $q$ values between 0 and 1. The $q$ value that corresponds to the minimum $\Sigma res^2$ is adopted as the most probable one. This value is then set as initial input parameter and is adjusted during the entire subsequent analysis, until the code converges to the final solution, which is considered as the best fit.
In our study, the $q$-search method will be used only when there is complete absence of radial velocity measurements.

The already existing 2-dimensional and 3-dimensional empirical relations \citep{Gazeas2009} will also be used as a first approximation of the absolute physical parameters of both components in each system (providing an uncertainty of about 10~per~cent) and will be tested, under the extreme conditions of stellar merger candidates.
When spectroscopy is not available or it is difficult to obtain on faint targets, these empirical relations will be used, to roughly estimate the absolute physical parameters of the systems.
These relations will be further calibrated and optimized in the frame of {\it CoBiToM Project} and they will eventually provide precise absolute physical parameters with uncertainty of less than 5~per~cent without the need of spectroscopy.

The orbital period of contact binaries present temporal variations, as a result of either mass transfer between their components, or mass loss via stellar wind or stellar aging through evolution processes. As a consequence, the
size of the orbit is varying, and this change could be detected by closely examining the observed times of minimum light residuals using the $(O-C)$ diagrams. Ground-based data as well as {\it Eclipse Time Variation (ETV)} data from the {\it Kepler} mission will be used for determination of the $dP/dt$, as well as for the search for tertiary components. In multiple systems, there are additional mechanisms that play an important role in shrinking and circularizing the orbits, and they are related to the combined action of Kozai cycles and tidal friction \citep{Kozai1962}. The determination of these parameters will be carried out using the time related $O-C$ diagrams, providing answers to whether the dynamical evolution of such systems is related to the Kozai cycles.

The stability zone for each binary will be determined in order to search for
the existence of exoplanets. This will be performed through simulations of test particles’ motion in the well-known {\it 3D circular restricted three-body problem (3D CR3BP)} \citep{Szebehely1967, Marchal2012}. The resulting stability zones will be associated with the evolution models of contact binaries. The possibility that the origin of some inflated hot Jupiters are linked to stellar mergers \citep{Martin2011} will be investigated through statistical analysis of physical and orbital properties.
The most crucial of them are the direction of the revolution compared to the parent star's rotation and the inflated radius of the planet.
The radius and spin will be estimated using a widely-used technique based on the {\it Rossiter–McLaughlin effect} \citep[RM;][]{Winn2005}, that implements radial velocity data and a hierarchical Bayesian flux-dependent model \citep{Thorngren2018,Sestovic2018}. In case of stellar host with spot activity, the orbital inclination and the direction of revolution of the exoplanets can be revealed using an innovative method called as {\it Transit Chord Correlation} \citep[TCC,][]{Dai2018}. This method has been already used with success to G and K stars with hot Jupiters.


\section{Notes on Individual Targets - Case Studies}

In this section, some case studies are presented as the first results of the {\it CoBiToM Project}.
These include two ultra-short period contact binaries, a gravitationally bound doubly eclipsing system, a contact binary with extremely low mass ratio, two cases of contact binaries that exhibit negative orbital period change and the case of a planet-host candidate.


\begin{figure}
\includegraphics[width=8.5cm,scale=1.0,angle=0]{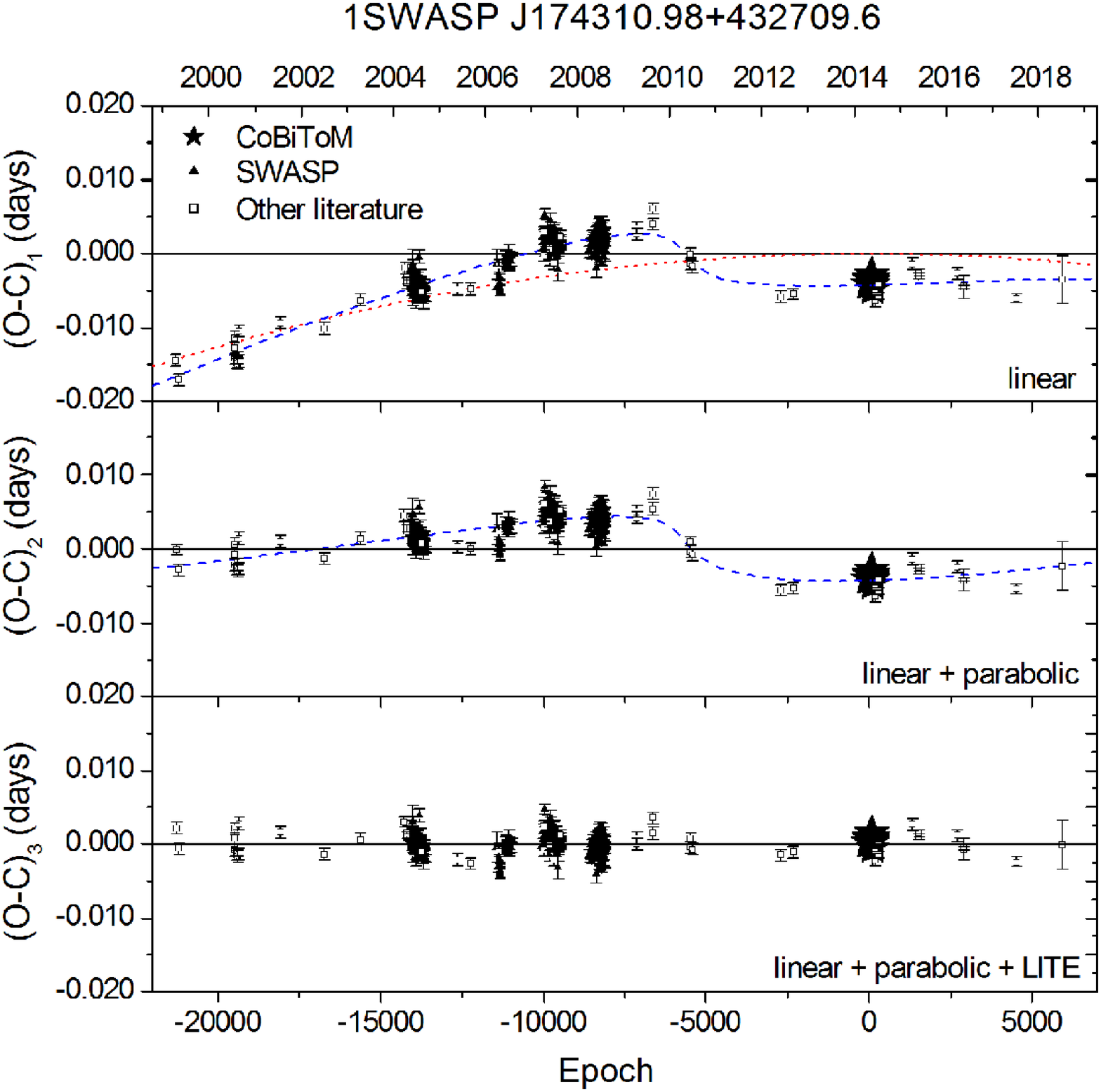}
\caption{Orbital period modulation analysis of the contact binary system 1SWASP~J174310.98+432709.6. Top: $(O-C)_{1}$ diagram, after subtracting the linear ephemeris and fitted by a sum of a parabolic and a LITE curves (blue dashed line). The red dotted line denotes only the parabolic term. Middle: $(O-C)_{2}$ diagram, after subtracting the parabolic and linear curves and fitted by a LITE curve.
Bottom: $(O-C)_{3}$ diagram, after the subtraction of all terms.}
\label{Fig2}
\end{figure}

\subsection{The orbital period modulation of 1SWASP~J174310.98+432709.6}

The orbital period modulation analysis is an important diagnostic tool for studying stellar mergers. It provides information about orbital period variations (secular and cyclic ones). Complex systems can exhibit orbital period variations, due to angular momentum loss via Kozai mechanism. A direct method of investigating the orbital shrinkage is through the $O-C$ analysis. A linear ephemeris usually describes the short-term orbit of contact binaries (i.e. for a couple of years). Times of minimum light can be used as an accurate tool for detecting negligible orbital period modulations.

\cite{Lohr2012} studied 53 ultra-short period binaries and found that only three of them (5.6~per~cent) exhibit rapid orbital period decrease. \cite{Molnar2019} found 108 systems with monotonic orbital period change rate in a sample of 22,462 (contact, near contact or detached) binaries in the OGLE sky survey \citep{Pietrukowicz2017}. This corresponds to about 0.5~per~cent of the entire sample. They also proposed that such a behavior may be caused by the Darwin tidal instability. \cite{Kubiak2006} detected period modulation in 134 contact binaries in a sample of 569 contact binaries, increasing significantly the percentage to 23.5~per~cent. The latest result is also confirmed by \citet{Lohr2015b}, who also found the low limit for
the higher-order multiplicity fraction among local galactic binaries to be around 24~per~cent.

Our investigation shows indications that the percentage of ultra-short period systems with noticeable orbital period change rate is much higher, and possibly exceeds 50~per~cent. The present study led to the conclusion that the system 1SWASP~J174310.98+432709.6 exhibits period decrease (Fig. \ref{Fig2}). It is clear that a linear ephemeris (Eq.~\ref{Eq1}) cannot explain the long-term trend in the $(O-C)_{1}$ residuals. In this equation, $T$ represents the time of observed minimum light, $T_{0}$ is the reference time (i.e. the time of one observed primary minimum), $P$ the orbital period and $E$ the Epoch. The polynomial regression for parabola on the $(O-C)_{1}$ values suggests a quadratic term, hence, the $(O-C)_{2}$ residuals can be described by Eq.~\ref{Eq2}, where $b$ is the quadratic coefficient linked to the orbital period change rate (b=$\frac{1}{2}\frac{dP}{dt}$P). 1SWASP~J174310.98+432709.6 presents a periodic modulation in its orbital period, which can be explained
by a tertiary component orbiting the contact binary. We have applied a quadratic term plus the light-time-effect (LITE) term \citep{Irwin1959} as seen in the Eq.~\ref{Eq3}, where $a_{12}$ is the projected semi-major axis, $i_{3}$ the orbital inclination of the tertiary component with respect to the system's orbital plane, $c$ the speed of light, $e_{3}$ the eccentricity of the orbit of the tertiary component, $\omega_{3}$ the longitude of periastron and $\nu$ is the true anomaly around the centre of mass of the triple system. The above methodology was applied in order to extract the orbital period modulation, utilizing the {\it LITE} software \citep{Zasche2009}, that uses statistical weight on the input eclipse timings according to the observation method used.

\begin{equation}
    \centering
      (O-C)_1 = T - (T_{0} + P \times E )
    \label{Eq1}
\end{equation}

\begin{equation}
    \centering
      (O-C)_2 = (O-C)_1 - b \times E^2
    \label{Eq2}
\end{equation}

\begin{equation}
    \centering
      (O-C)_3 = (O-C)_2 - \frac{a_{12}\sin i_3}{c} \left[\frac{1-e_{3}^2}{1+e_{3}\cos\nu}\sin(\nu+\omega_{3})+e_{3}\sin\omega_{3}\right]
    \label{Eq3}
\end{equation}

The orbital period change rate of the system 1SWASP~J174310.98+432709.6 was also studied by \cite{Lohr2012}, who found a value of $-6.37\times10^{-7}$~d~yr$^{-1}$, as listed in their Table 1.
The periodic modulation (due to a possible existence of a third component)
was not taken into consideration by the authors, as data from SWASP span for a very short period (approximately between 2004-2009). Gathering all available data from the literature, including our own from the {\it CoBiToM Project}, we found a period decrease rate of $-0.8963\times10^{-7}$~d~yr$^{-1}$ and a tertiary component with an orbital period of 19.1(4)~yr. The current technique proves that the study of $O-C$ diagrams is an excellent tool for investigating long-term orbital period modulations. 

\cite{Kubiak2006} studied several contact binaries from the OGLE database and showed that the majority of them present orbital modulation with $dP/dt$ evenly distributed between $\pm 2.3 \times 10^{-7}$~d~yr$^{-1}$, indicating that the decreasing period is not favored and the values are rather uniformly scattered.
In the same study, the authors found that 41.2~per~cent of the examined contact binary systems exhibit secular changes in their orbital period. This is a very high percentage, considering that there are no binary mergers observed, except the V1309~Sco red nova event.

It should be mentioned that the observed parabolic behavior could also reflect a longer term periodic variation caused by magnetic braking or by a third body, which can be clarified only after several years or decades of monitoring. This means that several $dP/dt$ values are biased from the existence of a tertiary components, which are likely to orbit contact binaries \citep{Pribulla2008,Deb2011,D'Angelo2006}.


\begin{figure*}
\includegraphics[width=8.0cm,scale=1.0,angle=0]{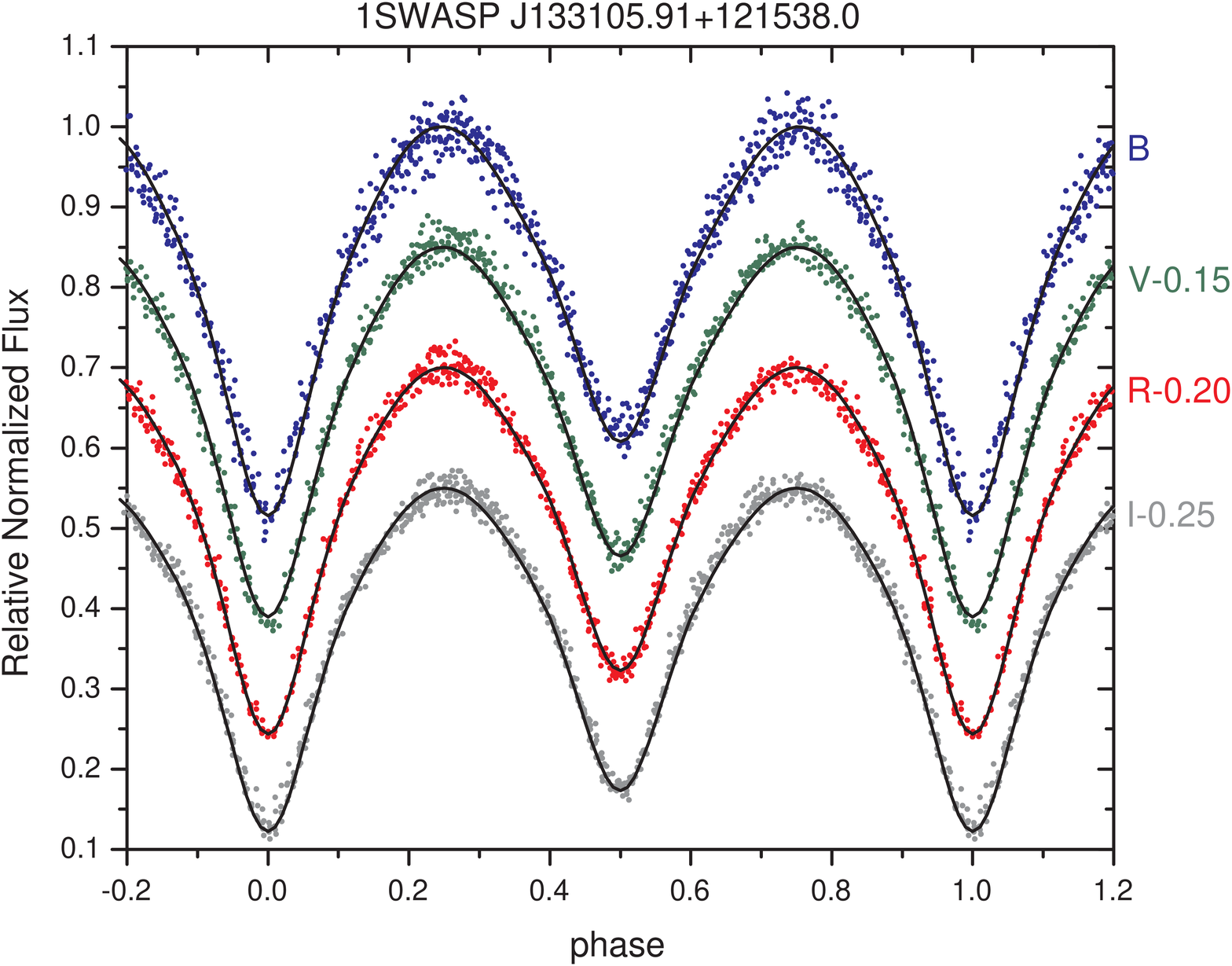}
\includegraphics[width=8.0cm,scale=1.0,angle=0]{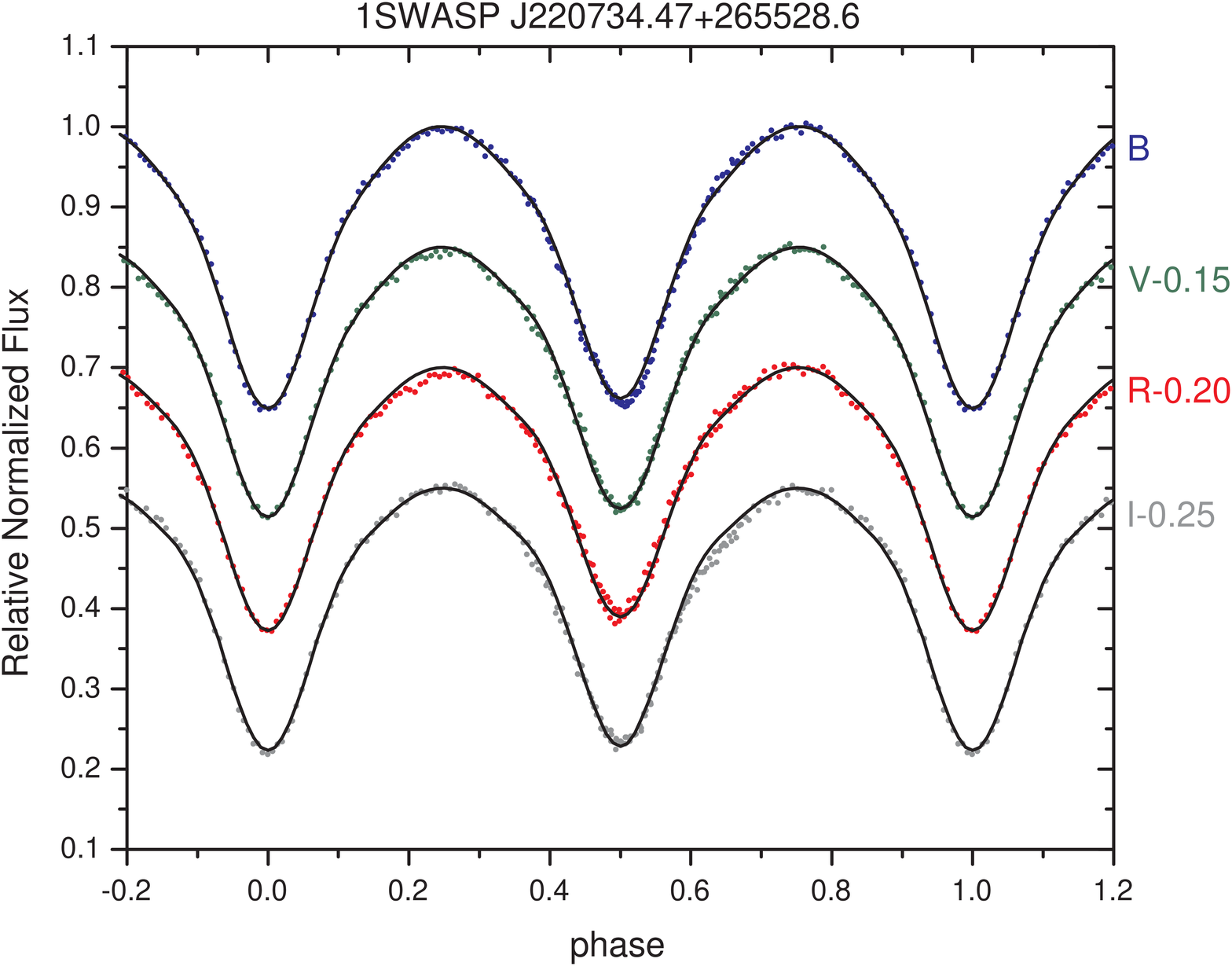}
 \caption{The observed light curves of the two ultra-short period contact binaries 1SWASP~J133105.91+121538.0 and 1SWASP~J220734.47+265528.6. We demonstrate a dense phase coverage in four photometric bands  that guarantees the monitoring of the brightness modulations, which is essential for the binary modelling.}
\label{Fig3}
\end{figure*}

\begin{figure}
\includegraphics[width=8.5cm,scale=1.0,angle=0]{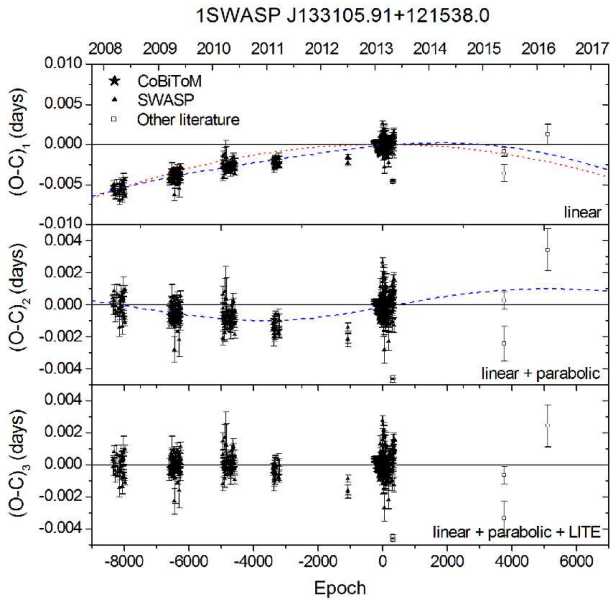}
\caption{The same as Fig. \ref{Fig2} but for the contact binary 1SWASP~J133105.91+121538.0.}
\label{Fig4}
\end{figure}

\subsection{The ultra-short period systems 1SWASP~J133105.91+121538.0 and 1SWASP~J220734.47+265528.6}

Examples of two ultra-short period binaries, studied in the frame of {\it CoBiToM Project}, are given in Fig. \ref{Fig3}. The systems are 1SWASP~J133105.91+121538.0 with $P_{\rm orb}=0.2180128(1)$~d, which was observed from UOAO in 2013, and 1SWASP~J220734.47+265528.6 with $P_{\rm orb}=0.2312354(1)$~d, which was observed from UOAO and Helmos Observatory in 2015.

The analysis of these systems showed that they belong to the group of contact binaries with extremely short orbital period. As a consequence, such systems have very small size, mass and revolution velocity, resulting in a very small angular momentum. Such systems lose mass and angular momentum via stellar wind.
Indeed, studies \citep[e.g.][]{Zola2006} show that the primary components of ultra-short period binaries are still Main Sequence dwarfs, while the secondary components are slightly above the TAMS line (they appear oversized and overluminous for their mass). The components of these systems have very small mass (0.5-1.0 $M_{\sun}$) and therefore their evolution is rather slow. According to \cite{Stepien2012}, such systems can survive more than 10-15~Gyr and the merging process is very slow. It is worth investigating the mass transfer rate, the orbital shrinking, and the mass exchange between their components, in order to confirm the above theory and study their merging scenarios.

Furthermore, the system 1SWASP~J133105.91+121538.0 was reported for orbital period modulation by \cite{Lohr2012}, who provided a value of
$-8.71\times10^{-7}$~d~yr$^{-1}$, as listed in their Table 1, based on SWASP photometric data. Our study includes all available data from the literature, together with our own (Fig. \ref{Fig4}). We found a period decrease rate of $-2.79(6) \times 10^{-7}$~d~yr$^{-1}$, that indicates orbital shrinkage, while the periodic modulation of the orbital period can be explained by a tertiary component orbiting the binary system with a period of $P_{3}$=19.1(4)~yr.

\begin{figure*}
\includegraphics[width=5.8cm,scale=1.0,angle=0]{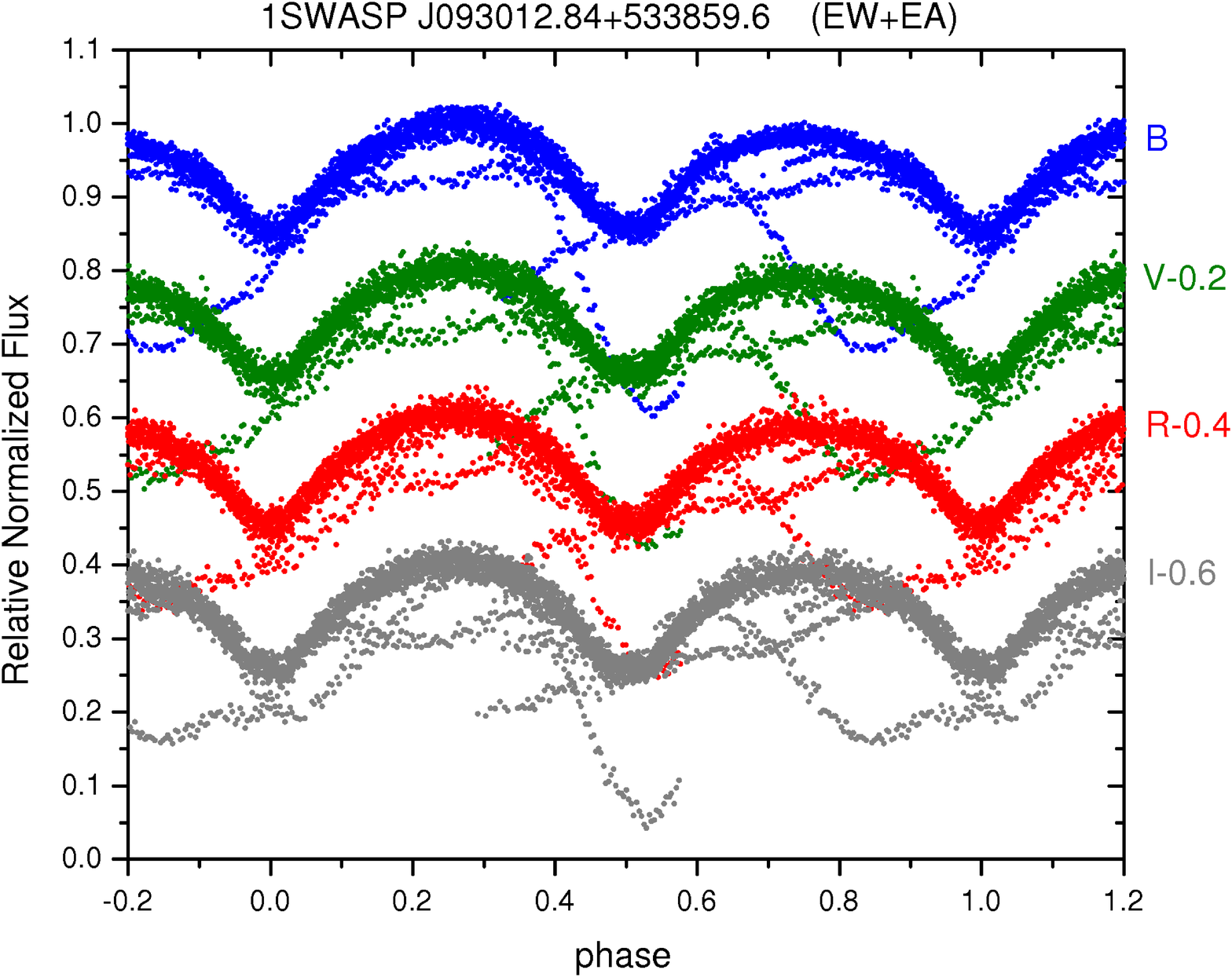}
\includegraphics[width=5.8cm,scale=1.0,angle=0]{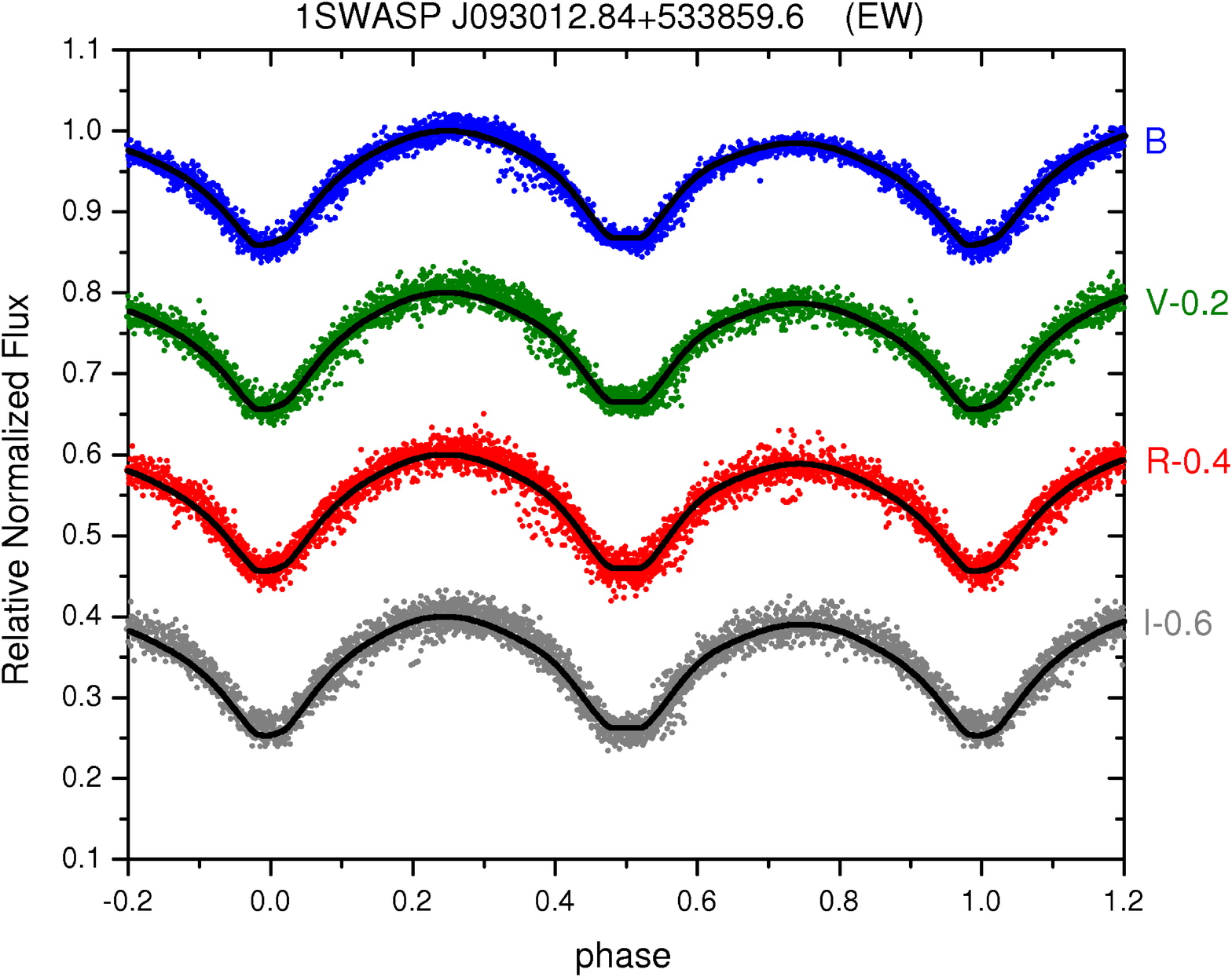}
\includegraphics[width=5.8cm,scale=1.0,angle=0]{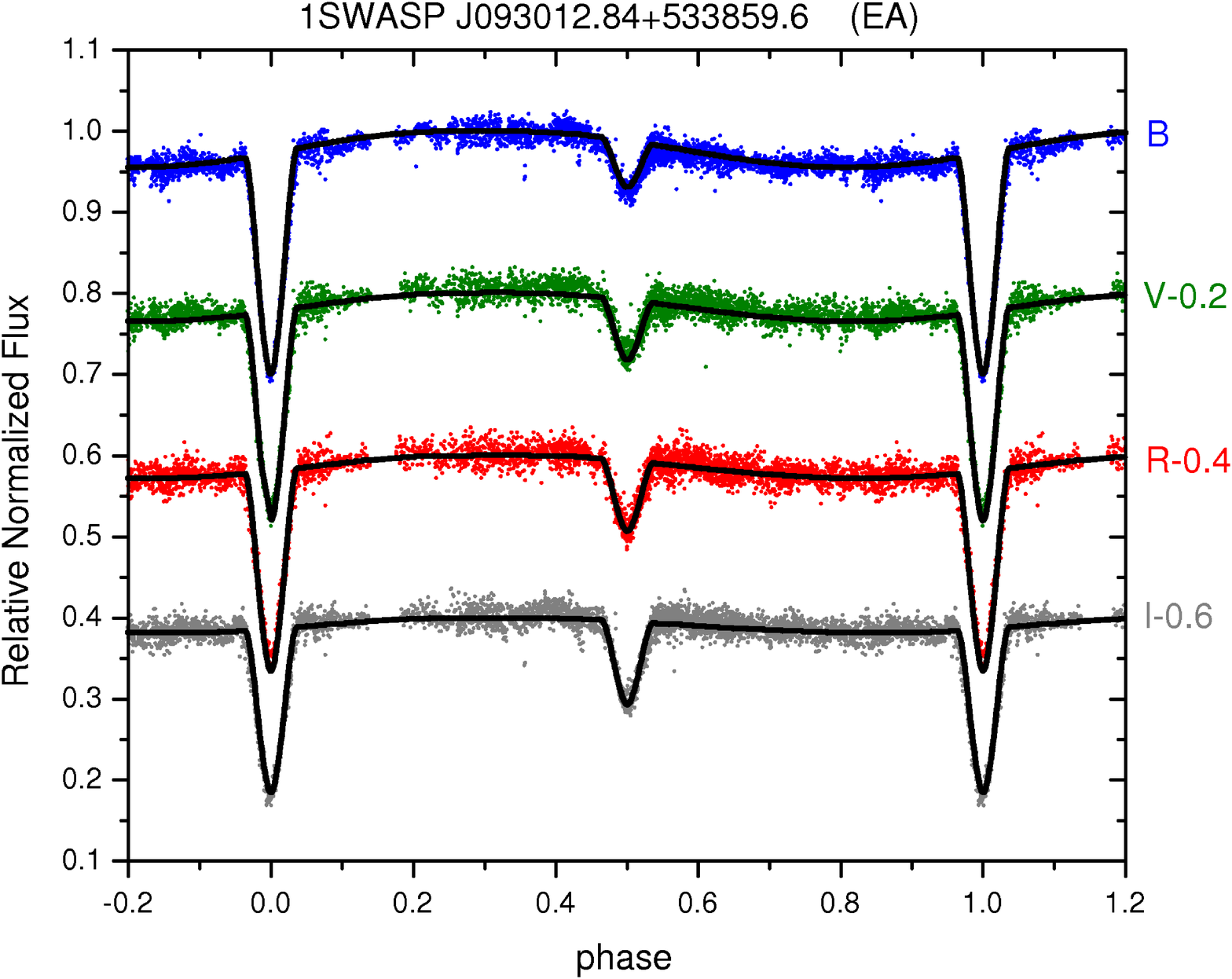}
\caption{The quadruple system 1SWASP~J093010.78+533859.5 presents a complicated light curve as a result of the combined eclipses within the quadruple system. The left panel shows the photometric data folded in the orbital period of the contact binary. After disentangling the contribution of the two binary systems, a contact and a detached binary are unveiled (central and right plots, respectively). Light curves of  both systems show asymmetries (the O' Connell effect), which are associated with cool spots on the photosphere of the primary (see text for details).
}
\label{Fig5}
\end{figure*}


\subsection{The quadruple system 1SWASP~J093010.78+533859.5}

This system was discovered in the frame of the SuperWASP sky survey \citep{Pollacco2006} and it was identified as a short-period eclipsing binary candidate by \cite{Norton2011}. Its orbital period was confirmed by \cite{Lohr2012} and it has been thoroughly studied ever since \citep{Lohr2013,Koen2014,Koo2014,Lohr2015a,Zasche2019}. The observed light curve seems to be a juxtaposition of a doubly eclipsing quadruple system, containing an ultra-short period and a detached binary (Fig. \ref{Fig5}, left panel) with an orbital period of 0.2277142(2)~d and 1.30550(4)~d, respectively. 1SWASP~J093010.78+533859.5 is also identified as a visual double star, easily resolved by large telescopes. The point sources with designation TYC~3807-759-1 and TYC~3807-759-2 are separated by 1.88~arcsec based on TYCHO catalog \citep{Hog2000}.

The gravitational bond between the two eclipsing systems is suggested by \citet{Zasche2019}, based on the eclipse timing variations of both binaries. The authors proposed a tight orbit between the two systems with a mutual orbital period of longer than 15~yr. Such an orbital period corresponds to an angular separation between 90-110 mas, taking into account the parallax measurement of $\sim$35~mas given by \citet{Agueros2009}. This angular separation, however, does not match the value given by TYCHO catalog.

\cite{Lohr2015a} obtained spectroscopic observations for the two easily resolved eclipsing systems and they confirmed the existence of five individual spectral signatures. They interpreted the fifth spectral signature as coming from a fifth stellar component nearby the detached eclipsing system, that seems to be “steady” and without indication of any orbital motion.

The  radial velocity of the center of mass for the EA system is found to be -12.3(2)~km~s$^{-1}$, being very close to the radial velocity of the fifth component, which is -11.5(1.5)~km~s$^{-1}$, and also to the EW system with -11.3(7)~km~s$^{-1}$. The above measurements coincide with each other, indicating that the fifth component forms a triple hierarchical system with the EA binary, while the EW binary is in close vicinity.

Astrometric parallaxes, given by the {\it Gaia}~DR2 \citep{Gaia2016,Gaia2018}, place the point sources TYC~3807-759-1 and TYC~3807-759-2 at a distance of $\sim$42.6(7)~pc and $\sim$69.85(1)~pc, respectively, indicating an impossible gravitational bond. The updated catalogue by {\it Gaia}~DR3 \citep{Gaia2020} does not include the distance of TYC~3807-759-1,  while the distance of TYC~3807-759-2 is updated with better accuracy at $\sim$70.114(3)~pc. Apparently, high order stellar multiplicity of TYC~3807-759-1 resulted in scattered parallax measurements, does not converge on a single distance estimation and makes the initial estimate of 42.6~pc questionable. On the contrary, the distance of $\sim$70~pc comes in agreement with the findings of both \citet{Koo2014} and \citet{Lohr2015a}, who also resulted in a similar distance range of 66(7)-77(9)~pc and 73(4)-78(3)~pc, respectively. At such a distance, and an angular separation of 1.88~arcsec, the two eclipsing systems have a minimum physical separation of $\sim$130~AU. This, together with the mass estimates of \citet{Koo2014}, implies a minimum orbital period >900~yr, which is not supported by the observations.

Magnetic activity could be the cause of eclipse timing variations, which are frequently observed in eclipsing binaries with cool components of K-spectral type. The $O-C$ diagram, presented by \citet{Zasche2019}, suggests a tight mutual interconnnection between the two systems, which seems to be unlikely, given the well defined angular separation. The cause of the periodic $O-C$ modulation could be a result of a combined magnetic activity in both eclipsing systems.
Further eclipse timing data are clearly needed, to resolve the origin of the 15~yr period.

Fig. \ref{Fig5} shows the original photometric data, folded with the orbital period of the contact binary (left panel). After disentangling the contribution of the two binary systems, the contact and the detached binaries are unveiled (central and right plot panels, respectively).
The appearance of the spot in different phases of the light curve on the detached system indicates spot migration, in agreement with the light curve presented by \cite{Koo2014} and \cite{Lohr2013}. This may suggest a non-synchronous rotation of the components in the detached system and an easier way to determine precisely the spin angular momentum.

Only two other doubly eclipsing quadruple systems, with at least one contact binary (EW-type) member, were known at the beginning of {\it CoBiToM Project}. Today, the number of similar cases has increased to 26, and they are presented in Table \ref{Table1}. Is is shown that most of these systems include detached (EA-type) eclipsing binaries in wide orbits. There are several doubly eclipsing systems still waiting to be confirmed as mutually bound and their light curves to be disentangled \citep{Zasche2019}.

\begin{table*}
\begin{center}
\caption{Contact binaries within doubly eclipsing quadruple systems. $P_{1}$ denotes the orbital period of the first eclipsing system (EW), while $P_{2}$ the orbital period of the second eclipsing system (EW or EA).}
\begin{tabular}{ lllll }
\hline
 system           & configuration       & $P_{1}$~(d)   & $P_{2}$~(d)   & references \\
\hline
1SWASP~J093010.78+533859.5	&(EW+EA)	& 0.2277135   & 1.30550     &\cite{Lohr2012}         \\
BW Dra + BV Dra 		    &(EW+EW)	& 0.29216     & 0.35007     &\cite{Batten1965}       \\
CoRoT 211659387             &(EW+EA)	& 0.393957    & 4.0005      &\cite{Hajdu2017}        \\
CY Tri (CzeV337+CzeV621)    &(EW+EA)	& 0.33343     & 0.53730     &\cite{Zasche2019}       \\
OGLE BLG-ECL-018877         &(EW+EA)	& 0.6008759   & 1.5565025   &\cite{Soszynski2016}    \\
OGLE BLG-ECL-061232         &(EW+EA)	& 0.3791298   & 1.4676043   &\cite{Soszynski2016}    \\
OGLE BLG-ECL-200747         &(EW+EA)	& 0.287215    & 42.7652100  &\cite{Soszynski2016}    \\
OGLE BLG-ECL-250817         &(EW+EA)	& 0.4396163   & 9.2515062   &\cite{Soszynski2016}    \\
OGLE BLG-ECL-197015         &(EW+EA)    & 0.3759299   & 6.53287     &\cite{Soszynski2016}    \\
OGLE BLG-ECL-100363         &(EW+EA)    & 0.5749267   & 4.3521616   &\cite{Soszynski2016}    \\
OGLE BLG-ECL-019637         &(EW+EW)    & 0.368949    & 0.4011300   &\cite{Soszynski2016}    \\
OGLE BLG-ECL-089724         &(EW+EA)    & 0.343487    & 3.4925576   &\cite{Soszynski2016}    \\
OGLE BLG-ECL-251128         &(EW+EW)    & 0.3786368   & 0.406083    &\cite{Soszynski2016}    \\
OGLE BLG-ECL-277539         &(EW+EA)    & 0.3753292   & 0.5779823   &\cite{Soszynski2016}    \\
OGLE BLG-ECL-282858         &(EW+EA)    & 0.3992092   & 0.539641    &\cite{Soszynski2016}    \\
OGLE BLG-ECL-352722         &(EW+EA)    & 0.5866713   & 3.28423     &\cite{Soszynski2016}    \\
OGLE BLG-ECL-406204         &(EW+EA)    & 0.5740634   & 1.7556268   &\cite{Soszynski2016}    \\
OGLE BLG-ECL-093829         &(EW+EA)    & 0.5210858   & 3.7452992   &\cite{Soszynski2016}    \\
OGLE BLG-ECL-104219         &(EW+EW)    & 0.457687    & 0.4683403   &\cite{Soszynski2016}    \\
OGLE BLG-ECL-165082         &(EW+EW)    & 0.9599463   & 1.092108    &\cite{Soszynski2016}    \\
OGLE GD-ECL-07157           &(EW+EA)	& 0.8128751   & 2.6694423   &\cite{Pietrukowicz2013} \\
OGLE GD-ECL-07443           &(EW+EA)	& 1.4512089   & 1.7501509   &\cite{Pietrukowicz2013} \\
OGLE GD-ECL-10263           &(EW+EW)	& 0.3787910   & 0.4208822   &\cite{Pietrukowicz2013} \\
OGLE LMC-ECL-16549		    &(EW+EA)	& 0.818033    & 164.79      &\cite{Graczyk2011}      \\
OGLE SMC-ECL-2715           &(EW+EA)	& 0.763215    & 1.02086     &\cite{Pawlak2013}       \\
OGLE SMC-ECL-4418           &(EW+EA)	& 0.71821     & 3.26509     &\cite{Pawlak2013}       \\
\hline
\end{tabular}
\label{Table1}
\end{center}
\end{table*}

\begin{figure*}
\includegraphics[width=8.0cm,scale=1.0,angle=0]{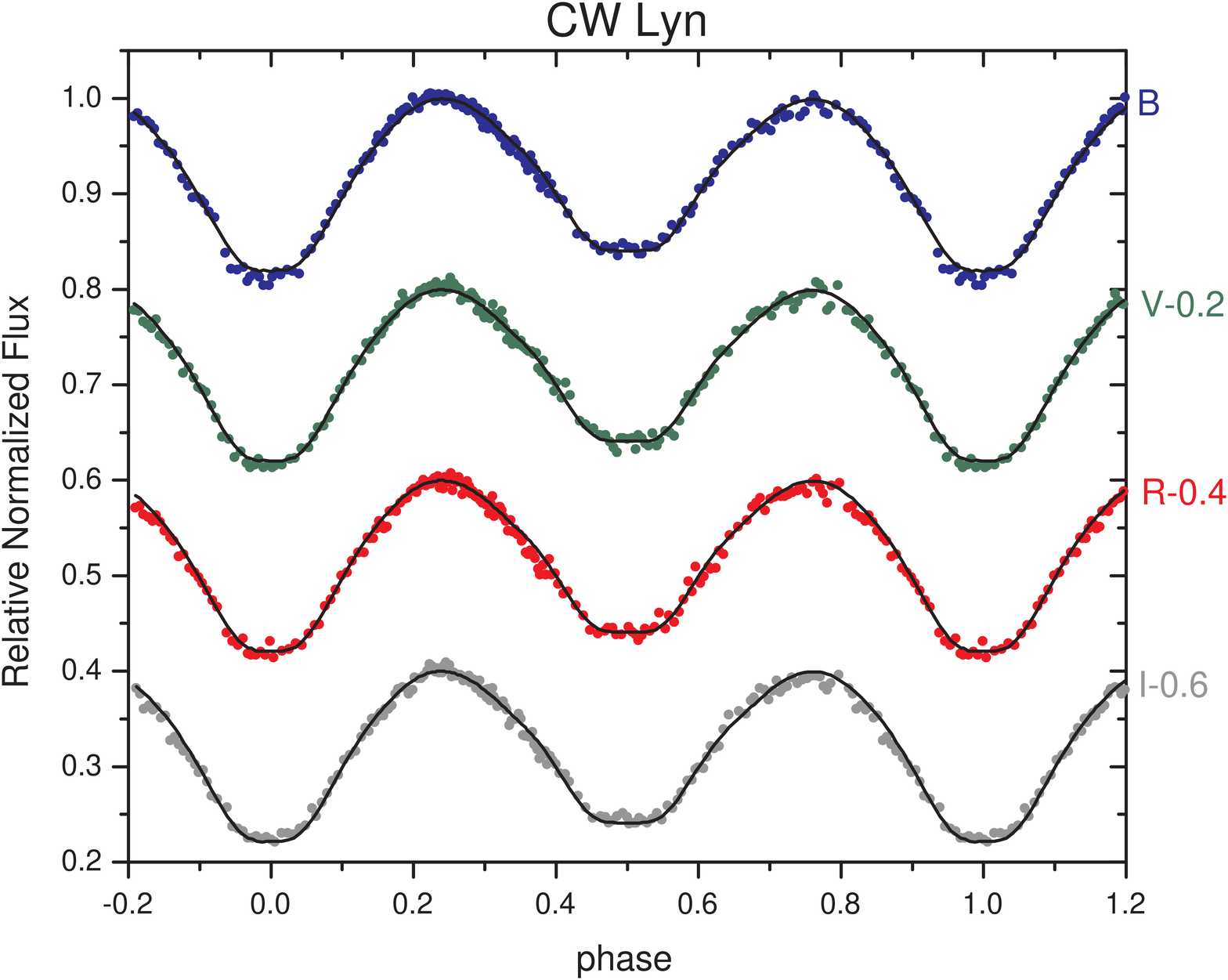}
\includegraphics[width=8.0cm,scale=1.0,angle=0]{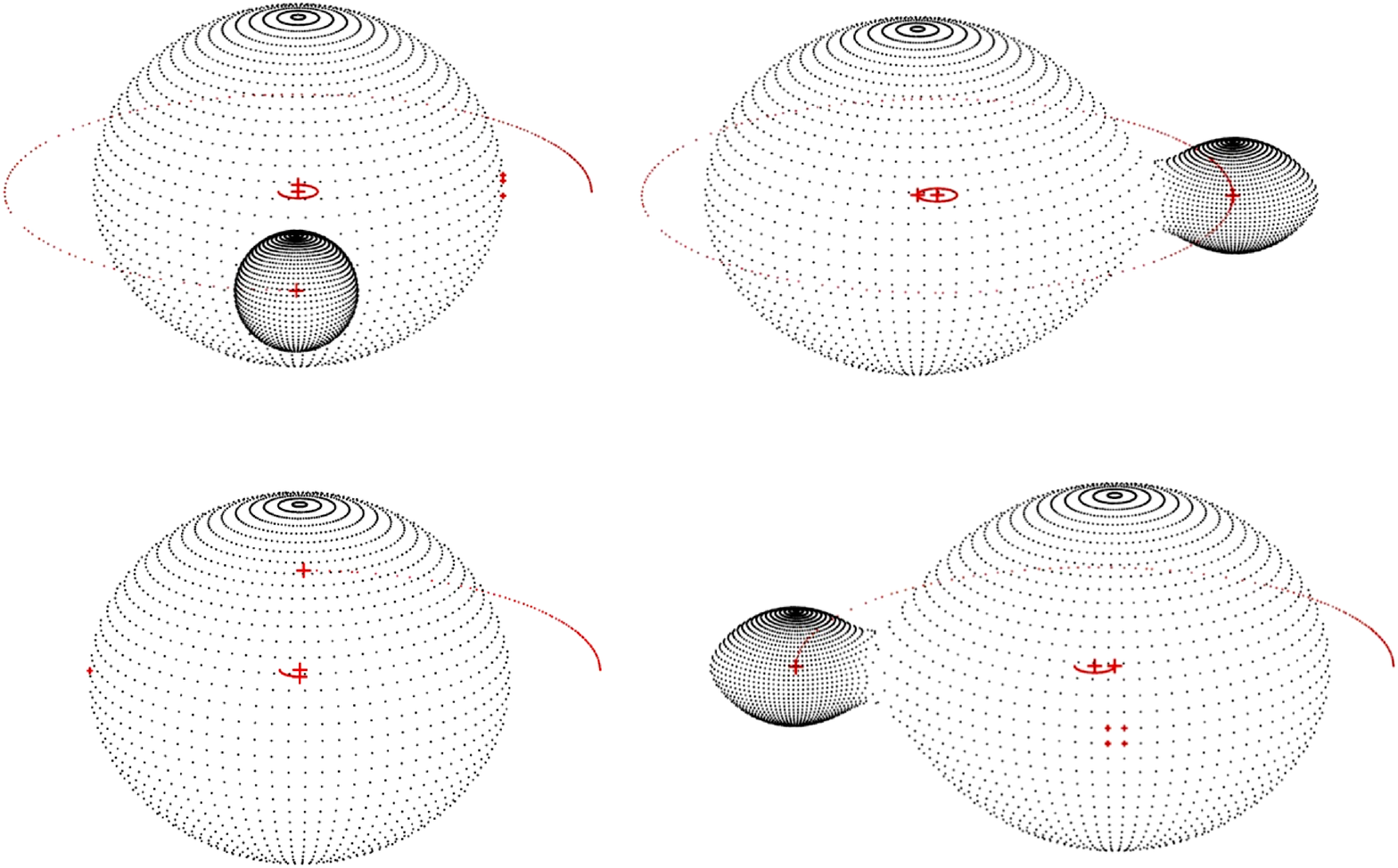}\\
\includegraphics[width=8.0cm,scale=1.0,angle=0]{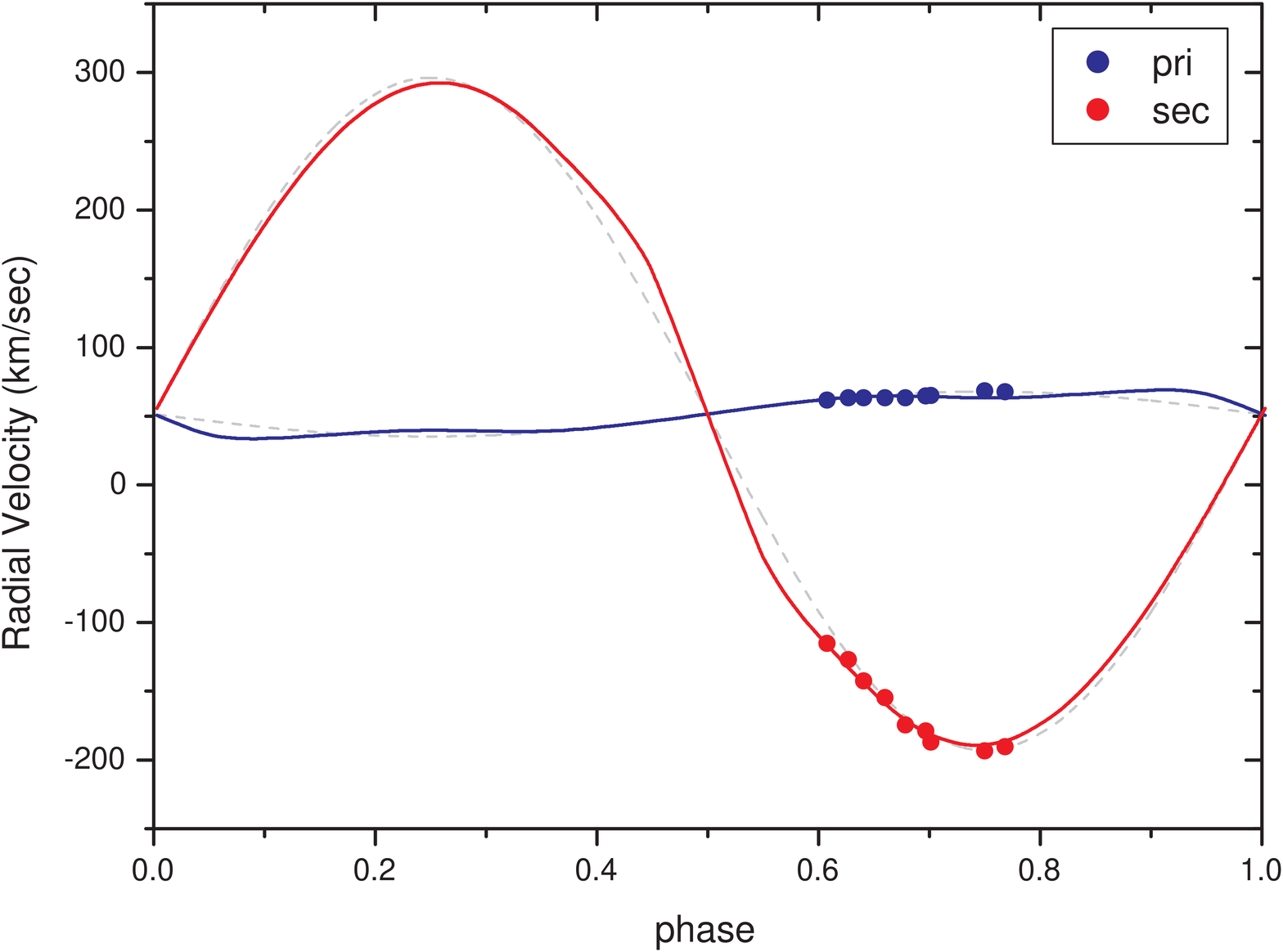}
\includegraphics[width=8.0cm,scale=1.0,angle=0]{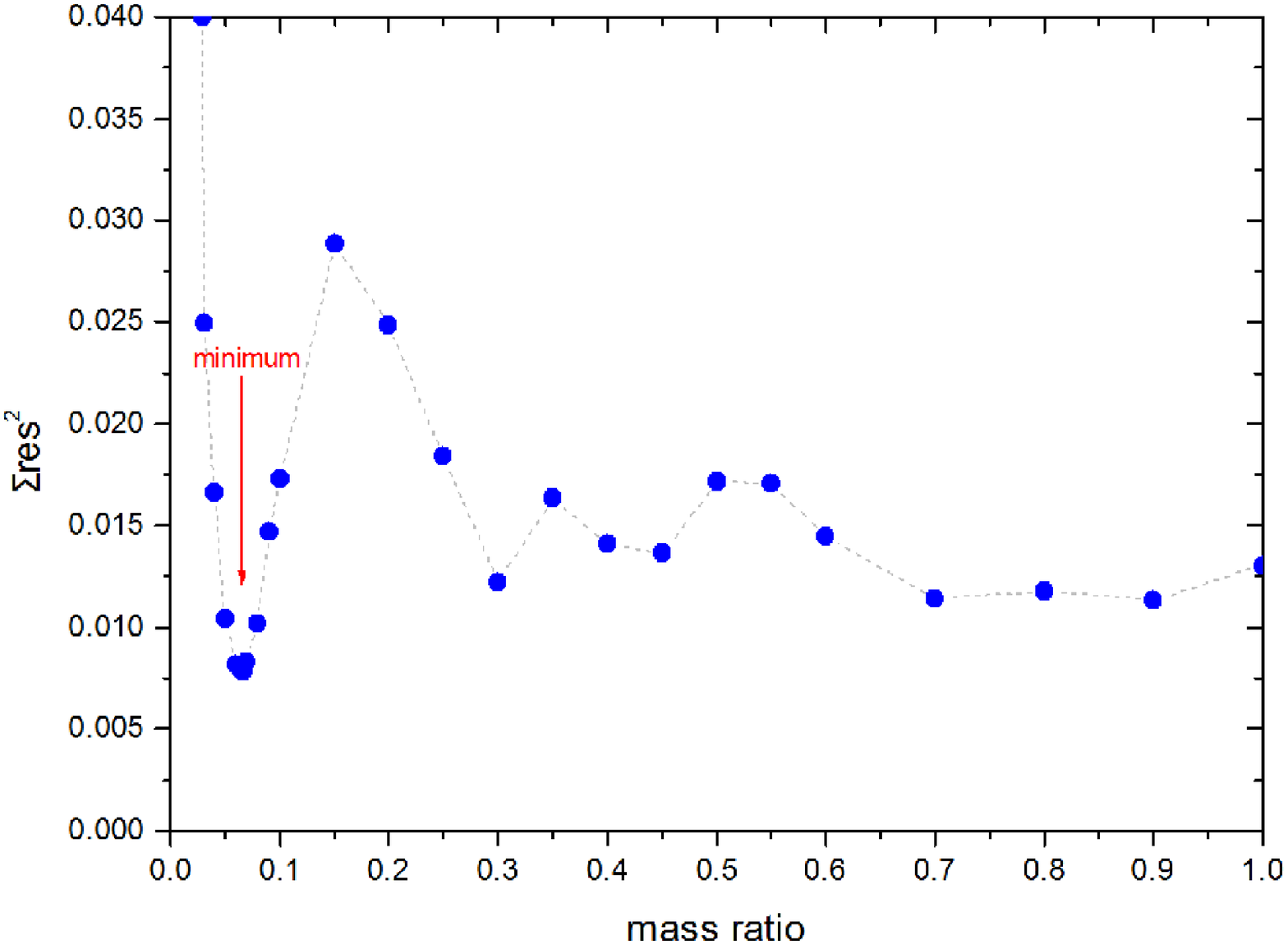}
\caption{The obtained $BVRI$ light curves of CW~Lyn (upper left) and the 3D model 
representation of the system (upper right). The radial velocities of \protect\cite{Pribulla2009} 
are fitted by our model parameters (bottom left), while the corresponding $q$-search 
method confirms the extreme value of the mass ratio value, as indicated in the bottom 
right panel with an arrow. The residuals in $q$-search method are extracted 
by comparing the photometric light curves with the theoretical model for the given 
mass ratio.}
\label{Fig6}
\end{figure*}


\subsection{The extreme mass ratio system CW~Lyn}

Information on low mass ratio systems is difficult to be found, since radial velocities (and spectroscopic observations in general) are by far more difficult to be obtained in comparison to the photometric ones, due to the faintness of the secondaries. There are a few targets under investigation in this regime. The systems V1187~Her \citep[$q=0.044$,][]{Caton2019}, V857~Her \citep[$q=0.065$,][]{Qian2005a,Qian2005b}, ASAS~J083241+2332.4 \citep[$q=0.068$,][]{Sriram2016} and  AW~UMa \citep[$q=0.099$,][]{Rucinski2015} are a few examples of this rare class of contact binaries, where the secondary component is extremely small in size and mass. \cite{Wadhwa2020} studied the contact binary system ZZ~PsA and found a very small mass ratio of 0.078(2) and an orbital period of 0.37405(3)~d. They concluded that the extreme mass ratio is accompanied by rapid mass transfer between the two components. They showed that ZZ~PsA is likely a highly unstable contact binary system and, hence, a potential merger (red nova) candidate.

The system CW~Lyn (HIP~42554) is observed in the frame of {\it CoBiToM Project} and is one more example of low-$q$ systems, identified via combined photometric and spectroscopic observations (Fig. \ref{Fig6}). It was discovered by the {\it HIPPARCOS} mission \citep{Perryman1997}. There is a limited number of studies for this system, namely the photometric study by \cite{Selam2004} and the spectroscopic study by \cite{Pribulla2009}. The spectral type is estimated to be F3/4V, which is in accordance with the 2MASS color index $(J-K)$=0.263~mag. The spectroscopic study resulted in no third light i.e. no tertiary component orbits the contact binary. It has a relatively long orbital period of 0.8124017(9)~d, while its $O-C$ diagram is rather flat, indicating no period modulation during the past two decades

The radial velocities of its components were poorly measured \citep{Pribulla2009} and therefore the mass ratio was not calculated. We performed a $q$-search on this system, and we resulted in a very low mass ratio of 0.067(2), close to the lower limit among the contact binaries known up to date (Fig. \ref{Fig6}, lower panels).

The mass ratio was also estimated with an alternative method. We used the WD code by utilizing Monte Carlo (MC) techniques, searching all the parameter space and the $q$ values between 0.01 and 1. This resulted in $q=0.066(2)$, adding confidence to our finding with the $q$-search method. Combining our findings together with the spectroscopic observations of \cite{Pribulla2009}, we conclude that the radial velocity amplitudes are $K_{1} = 16.4(1.2)$~km~s$^{-1}$ and  $K_{2} = 244.6(1.8)$~km~s$^{-1}$ and the systemic velocity is $V_{0} = 51.6(1.5)$~km~s$^{-1}$.

Such a configuration favours orbital shrinking due to Darwin instability \citep{Rasio1995}. Systems with very low mass ratio are expected to exhibit fast mass flow from the less massive (secondary) component towards the primary one. If such a scenario is true, there should be a systematic negative period change rate for these systems, indicating that these systems are driven fast towards merging. The system CW~Lyn does not show such a behavior, probably due to the lack of enough observations.


\subsection{The system KIC~9832227}

\begin{figure}
\includegraphics[width=8.5cm,scale=1.0,angle=0]{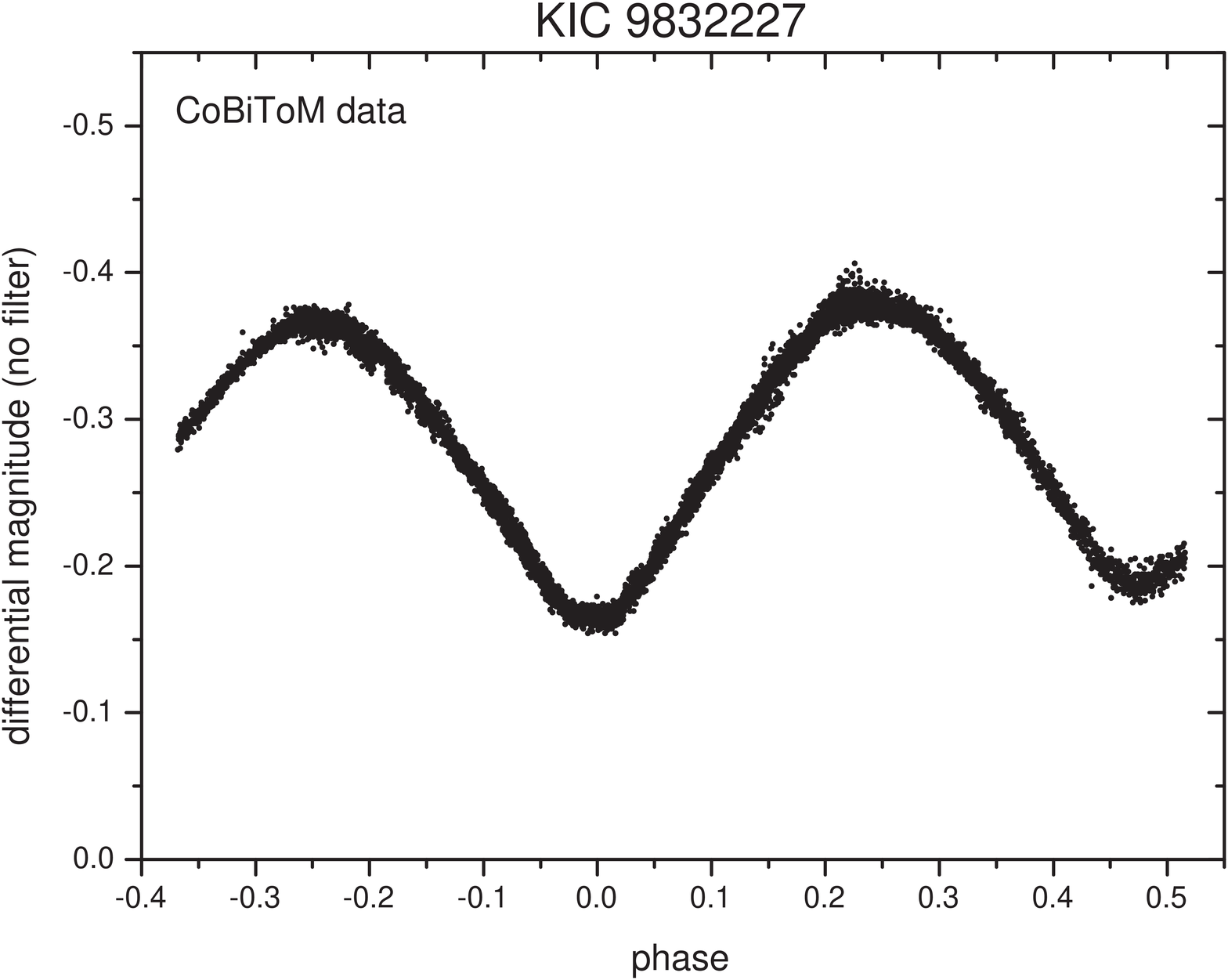}
\caption{The system KIC~9832227 presents a highly variable light curve.
The diagram above shows the unfiltered data obtained from Helmos Observatory.}
\label{Fig7}
\end{figure}

The system KIC~9832227 was discovered by \cite{Molnar2015} and it was also monitored in the frame of {\it CoBiToM Project}. It has a negative orbital period change rate and a highly variable light curve shape, indicating photospheric activity on its components. The O’ Connell effect is prominent and a single phase diagram cannot easily describe the photometric variability being gathered over a period longer than a few days (Fig. \ref{Fig7}).

The light curve presents shallow eclipses, indicating a low orbital inclination 51.19(10)$^\circ$ and period of 0.4579452(1)~d, which is typical for a contact binary. The most interesting feature, presented by \cite{Molnar2017}, is the negative period change rate, 
which was calculated by utilizing photometric data collected by {\it Kepler} mission and various ground-based observatories. The orbital shrinking was calculated to be so rapid, that KIC~9832227 was expected to become a red nova (and merge) within the next years (estimated to merge in 2022).

\cite{Socia2018} and \cite{Kovacs2019} spotted a misinterpretation in archival data by Molnar’s group and suggested that the system is not a red nova candidate. With a more robust and careful approach, the recent studies resulted in an contact configuration that includes a third component orbiting the contact binary, while the binary exhibits period decrease with a much lower rate ($-0.1097(47) \times 10^{-7}$~d~yr$^{-1}$).

\begin{figure}
\includegraphics[width=8.5cm,scale=1.0,angle=0]{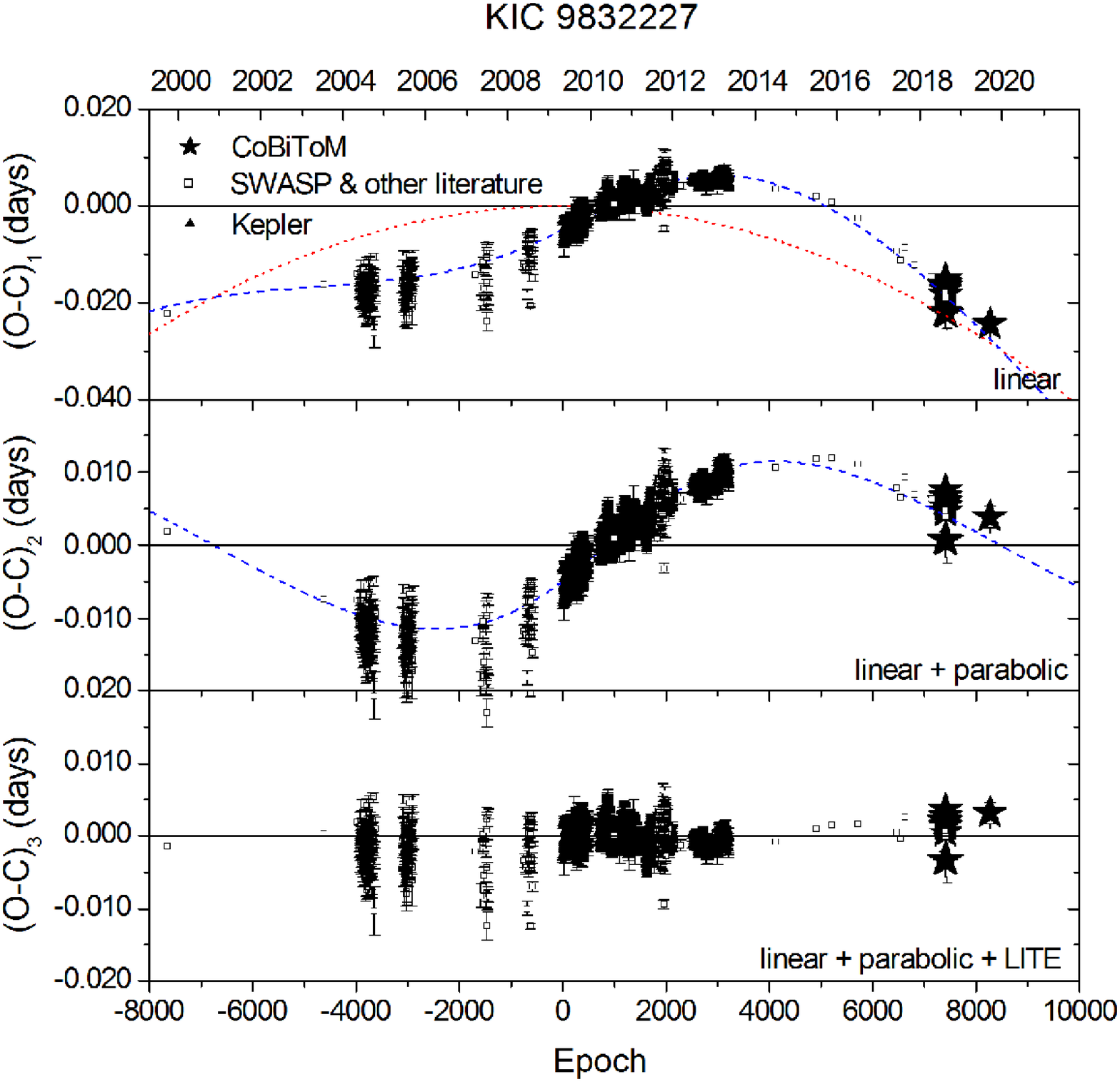}
\caption{The same as Fig. \ref{Fig2} but for the contact binary KIC~9832227.}
\label{Fig8}
\end{figure}

New times of minimum light are added in the $O-C$ diagram, based on observations conducted from UOAO and Helmos Observatory in Greece (Fig. \ref{Fig8}). It was found that the orbital period is decreasing with a rate of $-6.5766(11) \times 10^{-7}$~d~yr$^{-1}$, confirming the findings of \cite{Socia2018} and \cite{Kovacs2019}. With such a negative period change rate, it is not likely that the system will merge in the near future.

OGLE~299145 is a similar system presented also by \cite{Pietrukowicz2017}. It shows an orbital period change rate of $-1.7(2) \times 10^{-5}$~d~yr$^{-1}$, which is a rather high value. \cite{Tylenda2011} suggests that such a high value can be caused by tidal instability, that rapidly drives the systems towards shorter period values when the critical mass ratio is reached \citep{Rasio1995}. A similar merging process can also occur via fast mass loss through the Lagrangian point $L_2$. According to the same authors, such a process is believed to had occurred in V1309~Sco merger before the 2008 red nova outburst.

Systems similar to KIC~9832227 and  OGLE~299145 are currently monitored in the frame of {\it CoBiToM Project} in order to determine precisely the orbital period change and the possible red nova candidates in the forthcoming years.

\subsection{AW~UMa as a planet-host candidate}

At least 180 planetary candidates are presently known to be hosted in  binary or  multiple stellar systems \citep{Schwarz2016}. Nearly half of the exoplanets found within all types of binaries reside in those with very wide orbits and with average stellar separations greater than 1000~AU \citep{Kaib2013}.
The study of \citet{Horch2014}, by analyzing Kepler's data with the speckle imaging technique, suggests that the percentage of the binaries, that are also exoplanet hosts, is greater than what it was believed.
Specifically, half of them are probably wide binaries with separation ranging between 10 and 100~AU. The lack of exoplanets in close binaries still remains an open question \citep{Bonavita2020}.
Based on the exoplanet catalogue\footnote{\url{https://www.univie.ac.at/adg/schwarz/multiple.html}} of \cite{Schwarz2016}, the host binaries with small separation (<0.01~AU) have a white dwarf companion.
These systems, with $q$ ranging between 0.1 and 0.3, host gas giant planets, which have been detected with the {\it Transit Timing Variation} (TTV) technique, resembling the study of $O-C$ diagrams.
Despite the unfavorable dynamical conditions, there are evidences that planets can be formed in binary systems, even in close ones \citep{Bonavita2020}. However, there is no confirmed detection of exoplanet around a contact binary so far.

A possible candidate to host a circumbinary planet is AW~UMa. According to \cite{Pribulla2008}, this system might have a thick disk around the equatorial zone. However, according to \citet{Rucinski2015}, AW~UMa could be in detached and not in contact configuration.
RZ~Oph and V367~Cyg systems were also initially considered as contact binaries, however, further observations showed that they are probably in a semi-detached configuration, harbouring an accretion disk \citep{Zola1991, Zola2001}.

Planetary transits would be essential, if discovered in such systems, to confirm the
the above hypothesis. High precision photometric observations are necessary, something that can be obtained from space-borne observatories. AW~UMa was observed with MOST satellite, but no transits were detected during the observing windows \citep{Rucinski2013}. \cite{Elkhateeb2014} estimated the period change rate of AW~UMa 
($-4.1 \times 10^{-7}$~d~yr$^{-1}$), suggesting a relatively fast orbital shrinkage. They did not notice a negligible time modulation that is visible on the $O-C$ residuals and, therefore, the authors did not consider any tertiary component.

\begin{figure}
\includegraphics[width=8.5cm,scale=1.0,angle=0]{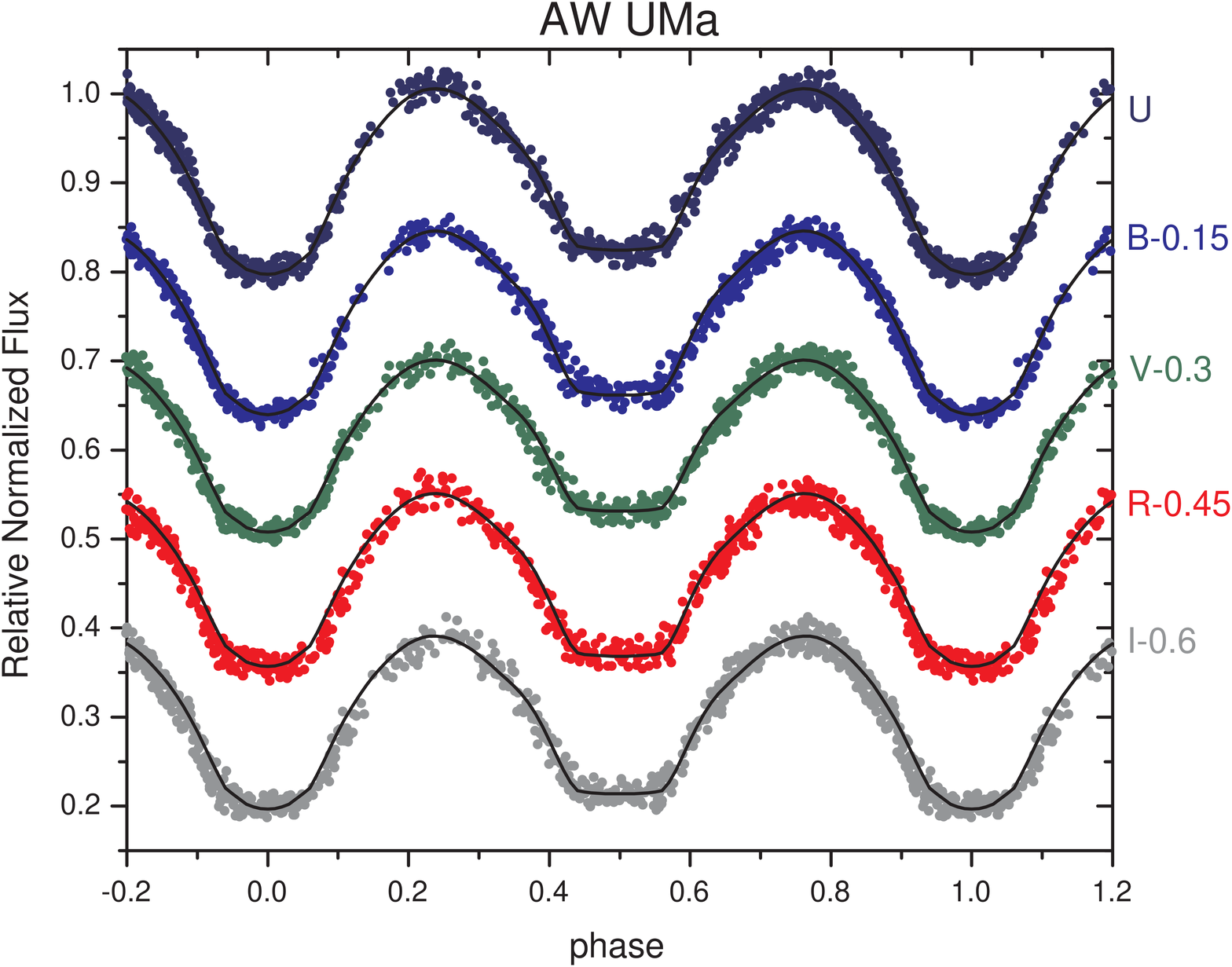}
\caption{The system AW~UMa has a very small mass ratio (0.099) and a high orbital inclination (81.29$^\circ$). As a result, the secondary component is totally eclipsed at phase=0.50. Our photometric light curves in $UBVRI$ bands are shown above. In this plot, the theoretical model of a contact system without a circumbinary disk is shown together with the observed data.
}
\label{Fig9}
\end{figure}

\begin{figure}
\includegraphics[width=8.5cm,scale=1.0,angle=0]{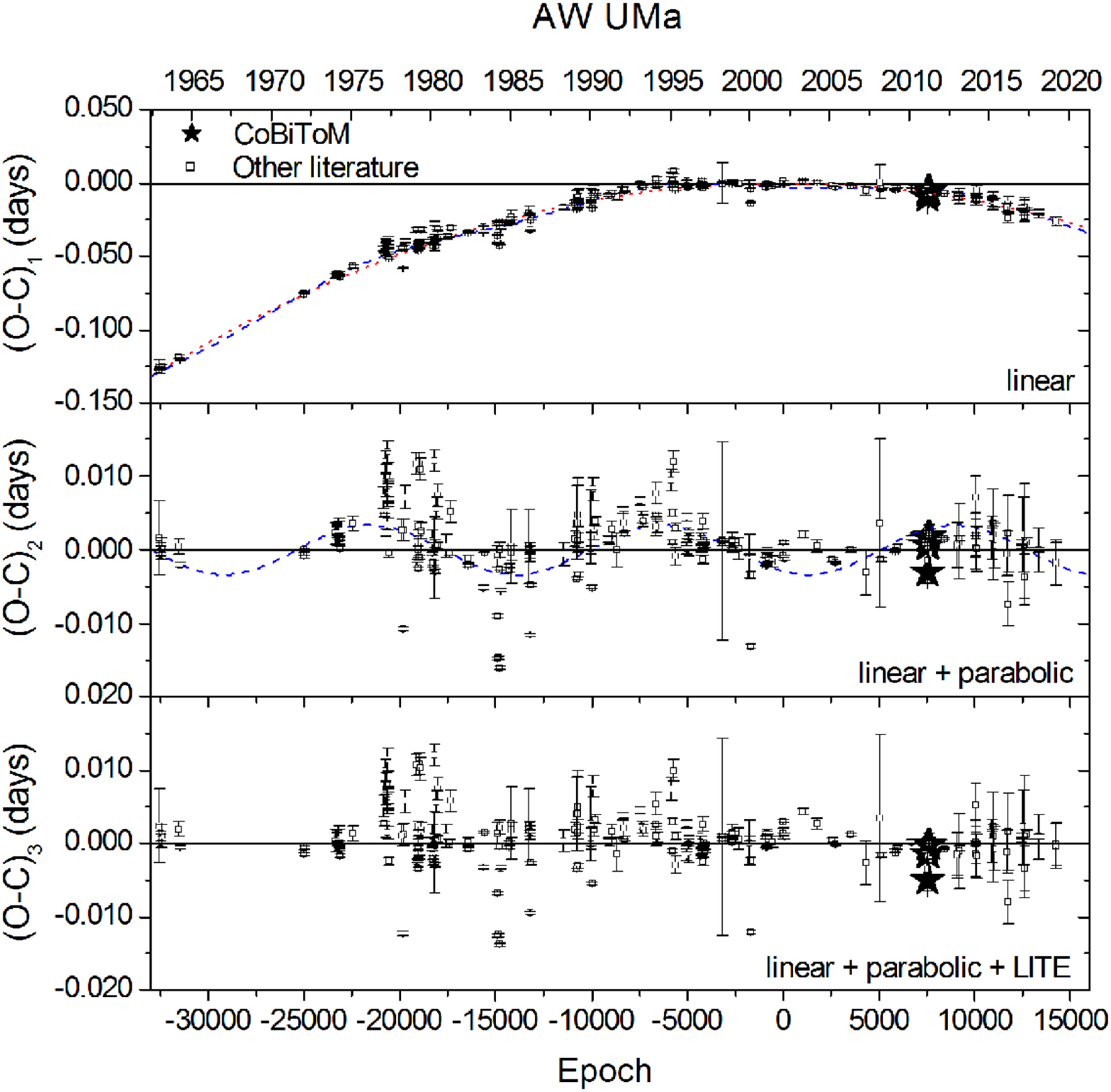}
\caption{The same as Fig. \ref{Fig2} but for the contact binary AW~UMa.}
\label{Fig10}
\end{figure}

AW~UMa has been observed from UOAO (Fig. \ref{Fig9}) in order to continue the long-term monitoring of its period modulation. Combining the available times of minima from the literature together with our own, we analyzed the $O-C$ diagram for this rare contact binary (Fig. \ref{Fig10}). Our study resulted in an orbital period change rate of approximately $-2 \times 10^{-7}$~d~yr$^{-1}$ for the contact binary and a cyclic period modulation, indicating a periodic magnetic activity or the existence of a third star orbiting the binary with a period of 18.3(4)~yr.

Detached systems are also candidates to host a giant planet (or inflated hot Jupiter), such as AV~CMi \citep{Liakos2010,Liakos2012b}. The authors found a third body transiting one of the two components of the binary system. Their models resulted in a brown dwarf stellar body or even a giant planet with a radius ranging between 4.4-6.4~$R_{\rm Jup}$, being a candidate of an inflated hot Jupiter orbiting one of the two components. Such systems can be detected
in the space-borne data of detached eclipsing systems \citep{Maxted2016}.

Nevertheless, the abundance and properties of contact binaries and low-mass detached binaries are consistent with being possible progenitors of a primordial planetary systems. Several Jupiter-mass planets can be formed in a massive compact disk that is created after a merging event. Gravitational scattering between them can explain the high incidence of eccentric, inclined, and retrograde orbits \citep[][and references therein]{Dawson2018}. The nature and abundance of W~UMa-type systems make them a plausible candidate population for hosting hot Jupiters. Systems that are surrounded by a circumbinary disk can potentially create the circumstances for planetary formation.

Tertiary components discovered through the $O-C$ studies can provide an insight on the mass of circumbinary objects. The detection/existence of a hot Jupiter orbiting a contact binary or a fast rotator will help in further understanding of the merging process.

\begin{table*}
\caption{Physical and orbital parameters of the systems studied in this present work, shorted in ascending orbital period. $P_{\rm orb}$ represents the orbital period of the contact binary, while $P_{3}$ represents the orbital period of the additional component orbiting the contact binary. The standard error for each value is given in parentheses, and corresponds to the last decimal places.}
\begin{flushleft}
\begin{tabular}{lcccccccc}
\hline
system &	$P_{\rm orb}$	& $M_1~(M_{\sun})$ & $M_{2}~(M_{\sun})$ &	$q$	&  multiplicity	&	$dP/dt$~($\times 10^{-7}$~d~yr$^{-1}$)	& $P_{3}$~(yr) & ref. \\
\hline											
1SWASP~J093010.78+533859.5	&	0.2277142(2) 	    & 0.860(20) & 0.341(11) &	0.397(6)\textsuperscript{sp}	&	yes	&	-	        & >15        & 1,2,5   \\
1SWASP~J133105.91+121538.0	&	0.2180128(1) 		& 0.789(4)  & 0.550(16) &	0.70(5)\textsuperscript{ph}	&	yes	&	-2.79(6)	& 11(6)      & 5       \\
1SWASP~J174310.98+432709.6	&	0.2581081(1) 	    & 0.937(24) & 0.345(15) &	0.37(5)\textsuperscript{ph}	&	yes	&	-0.8963(8)	& 19(1)      & 5       \\
1SWASP~J220734.47+265528.6	&	0.2312354(1) 		& 0.824(1)  & 0.569(14) &	0.69(5)\textsuperscript{ph}	&	no	&	-	        & -          & 5       \\
AW~UMa			            &	0.4387254(2) 	    & 1.336(12) & 0.132(11) &	0.099(3)\textsuperscript{sp}	&	yes	&	-2.01204(4) & 18.3(4)    & 3,5     \\
CW~Lyn			            &	0.8124017(3) 	    & 1.707(5)  & 0.114(2) &	0.067(2)\textsuperscript{sp}	&	no	&	-	        & -          & 5       \\
KIC~9832227		            &	0.4579452(1) 	    & 1.395(11) & 0.318(5) &	0.228(3)\textsuperscript{sp}	&	yes	&	-6.5766(11)	& 19.1(4)    & 4,5     \\
\hline
\end{tabular}
\begin{small}
sp: spectroscopically determined, ph: photometricaly determined \\
references:  (1): \cite{Zasche2019}, (2): \cite{Lohr2015a}, (3): \cite{Rucinski2015}, (4): \cite{Molnar2017}, (5): present work
\end{small}
\end{flushleft}
\label{Table_Results}
\end{table*}

\section{Summary and Discussion}

The present work is the `kick-off' of the {\it CoBiToM Project} where the first results of some selected contact binaries towards merging are presented.

It highlights the aims of the project and presents examples of the multi-method approach in investigating stellar evolution towards merging.
The systems presented in this study fulfill the criteria of a pre-merging period and exhibit a `diverge' behavior. Some of them show a combination of more than one criteria, making them even more interesting objects for further studies.
Several factors influence the rate, at which a particular system evolves towards coalescence. The most critical ones appear to be: a) The relationship between orbital and spin angular momentum \citep{Rasio1995}, b) the degree of contact \citep{Rasio1995a}, and c) the angular momentum loss \citep{Stepien2012}. The results of this study are summarized in Table \ref{Table_Results} and the major findings are the discussed in the following paragraphs.

\begin{figure}
\includegraphics[width=8.5cm,scale=1.0,angle=0]{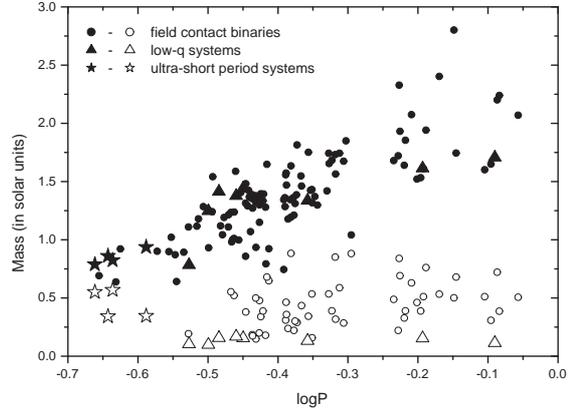}
 \caption{Mass distribution of contact binary components with respect to the orbital period, as presented in the sample of 138 contact binaries \citep{Gazeas2021}. Filled circles represent the primary and empty circles the secondary components. Triangles represent the low mass ratio systems, found in the same sample. Ultra-short period binaries are denoted with asterisks at the lower-left part of the diagram. For all systems in this sample the mass ratio was  determined spectroscopically.}
\label{Fig11}
\end{figure}

The mass distribution of primary and secondary components in the sample of 138 contact binaries \citep{Gazeas2021} as a function of the orbital period is shown in Fig. \ref{Fig11}. The derived parameters of this sample are extracted through a combined spectroscopic and photometric analysis. In this plot it is obvious that ultra-short period binaries consist of low-mass (and apparently very small in size) components. Their contact configuration indicates that the components are very close to each other, a fact that is reflected to a) the ultra-short orbital period, b) the very small mass and radius, and c) the eventually very low angular momentum. It is found that the four ultra-short period systems presented in this study do not have extreme mass ratios.
The ultra-short orbital period systems 1SWASP~J133105.91+121538.0 and 1SWASP~J174310.98+432709.6 show both cyclic orbital period modulation and a negative trend in their $O-C$ diagrams. The latter findings suggest orbital shrinkage and tertiary bodies orbiting around the contact binaries to be the most possible explanation. The $O-C$ diagram of 1SWASP~J220734.47+265528.6 shows neither a parabolic trend, nor any other apparent orbital period modulation, being in agreement with the earlier study by \citet{Lohr2012}. The investigation of ultra-short period binaries and their orbital period modulation in the frame of {\it CoBiToM Project} will bring out more merger candidates.

Low-$q$ systems seem to have longer orbital period (>0.3~d) and span across the entire range of the orbital period values. CW~Lyn is a known contact binary that has been poorly studied. In this work, we found an extremely small mass ratio value of 0.067(2). The lack of long-term photometric monitoring of this target does not allow for a conclusive result on its orbital period modulation.
The low mass ratio of AW~UMa and CW~Lyn suggests a very unstable evolution state for these systems that fulfill the Darwin instability criteria \citep{Li2006}. Our study also indicates the possible existence of a tertiary component orbiting AW~UMa, something that has not been mentioned by any study up to date.

{\it CoBiToM Project} provides already promising results, like the fact that KIC~9832227, 1SWASP~J133105.91+121538.0, and 1SWASP~J174310.98+432709.6 exhibit orbital period decrease and that tertiary components are possibly orbiting around them. These findings were achieved by long-term monitoring and dedicated observations with specific telescopes, as well as a combined data reduction and the multi-method approach.
The majority of systems examined in this study (five out of seven) evolve within a multiple stellar environment. Some of them are known to be members of quadruple systems, some others were found to be triple systems
by means of spectroscopic observations, while other systems unveiled the existence of additional components through the $O-C$ analysis.
The latter method proved to be the most precise and efficient on orbital evolution and multiplicity detection.

A statistical study of a large number of merger candidates is essential while seeking evolutionary trends. Well calibrated empirical relations across the entire range of the parameters will be a strong tool for calculating the physical properties of contact binary stars, especially in the absence of spectroscopic data. Systems that diverge from the `typical' behavior in any way will be a subject under detailed study.

With this work we emphasize the scientific impact of dedicated contact binary research in Stellar Astrophysics and we present the first results of a wide range of contact binaries, as they appear in various astrophysical environments: field targets, members of multiple systems or clusters, ultra-short period binaries, systems with fast evolution towards merging or planet-host candidates. The observing strategy aims to provide multi-band ground-based photometry for all targets of interest, radial velocity measurements, and follow up observations for those collected from space missions. The corollary of such an observing project is the ability to monitor a statistically large sample (n > 100) of contact binaries that have been found to exhibit extraordinary behavior and study their peculiarities in detail. It is essential for stellar models to be based on high quality photometric and spectroscopic data, in order to assess the outcome of the project.
The outcome of such a research could potentially revise the textbooks in the field of stellar evolution and planetary formation \citep{beccari_boffin_2019}, making its scientific impact and value significant. {\it CoBiToM Project} will provide a strong and solid observational basis, contributing to a robust and consistent astrophysical theory.




\section*{Acknowledgements}
This research is co-financed by Greece and the European Union (European Social Fund-ESF) 
through the Operational Programme ``Human Resources Development, Education and Lifelong Learning'' 
in the context of the project 
Strengthening Human Resources Research Potential via Doctorate Research 
(MIS-5000432), implemented by the State Scholarships Foundation (IKY).
This work utilizes data from the robotic and remotely controlled telescope of the 
University of Athens Observatory (UOAO), located at the National and Kapodistrian University 
of Athens, Greece. Part of this work is also based on observations obtained with the 
1.2~m Kryoneri telescope, located at Corinthia, Greece and the 2.3~m Aristarchos 
telescope, located at Helmos Observatory, Achaia, Greece. Both telescopes are operated 
by the Institute for Astronomy, Astrophysics, Space Applications and Remote Sensing of the 
National Observatory of Athens.
The authors wish to thank the collaborators at the observing facilities for their support 
and the telescope time allocation.
The authors wish to thank Prof. A. Norton who reviewed the current manuscript and gave 
valuable comments that improved our work.


\section*{Data Availability}
The data underlying this article are available upon request to the corresponding author.


\bibliographystyle{mnras}
\bibliography{biblio}
\bsp	
\label{lastpage}
\end{document}